\documentclass[%
 reprint,
superscriptaddress,
 amsmath,amssymb,
 aps,prx
]{revtex4-2}

\usepackage{graphicx}
\usepackage{dcolumn}
\usepackage{bm}
\usepackage{svg}
\usepackage{microtype} 
\usepackage{paralist}
\usepackage[normalem]{ulem}
\usepackage[unicode=true, colorlinks=true, citecolor={blue!80!black}, urlcolor={blue!50!black}, linkcolor = {blue!80!black}]{hyperref} 
\usepackage{array}
\usepackage{siunitx}
\usepackage{wrapfig}
\usepackage{enumitem}
\usepackage{amsmath,amssymb}
\usepackage{array}
\usepackage{soul}
\usepackage{centernot}
\usepackage{bibunits}
\usepackage[titletoc, title]{appendix}
\usepackage{titletoc}
\usepackage{lineno}

\newcommand*{\balancecolsandclearpage}{%
  \cleardoublepage
  \twocolumngrid
}

\usepackage{tikz}

\newcommand{\tikzcmark}{%
\tikz[scale=0.23] {
    \draw[line width=0.7,line cap=round] (0.25,0) to [bend left=10] (1,1);
    \draw[line width=0.8,line cap=round] (0,0.35) to [bend right=1] (0.23,0);
}}

\newcommand{\ket}[1]{\left| #1 \right>}
\newcommand{\bra}[1]{\left< #1 \right|}
\newcommand{\ketbra}[1]{\left| #1 \right>\left< #1 \right|}

\newcommand{\expo}[1]{\text{e}^{ #1 }}

\newcommand{\sx}{\hat{\sigma}_x}
\newcommand{\sy}{\hat{\sigma}_y}
\newcommand{\sz}{\hat{\sigma}_z}
\newcommand{\sm}{\hat{\sigma}_-}

\newcommand{\ha}{\hat{a}}
\newcommand{\had}{\hat{a}^\dagger}
\newcommand{\hH}{\hat{H}}
\newcommand{\hPi}{\hat{\Pi}}
\newcommand{\hn}{\hat{n}}
\newcommand{\hU}{\hat{U}}

\begin{document}

\preprint{APS/123-cQED}

\title{Synthetic high angular momentum spin dynamics in a microwave oscillator}

\author{Saswata Roy}
\affiliation{Department of Physics, Cornell University, Ithaca, NY, 14853, USA}

\author{Alen Senanian}
\affiliation{Department of Physics, Cornell University, Ithaca, NY, 14853, USA}

\author{Christopher S. Wang}
\affiliation{James Franck Institute and Department of Physics, University of Chicago, Chicago, IL, 60637, USA}

\author{Owen C. Wetherbee}
\affiliation{Department of Physics, Cornell University, Ithaca, NY, 14853, USA}

\author{Luojia Zhang}
\affiliation{Department of Physics, Cornell University, Ithaca, NY, 14853, USA}

\author{B. Cole}
\affiliation{Department of Physics, Syracuse University, Syracuse, New York 13244-1130 USA}

\author{C. P. Larson}
\affiliation{Department of Physics, Syracuse University, Syracuse, New York 13244-1130 USA}

\author{E. Yelton}
\affiliation{Department of Physics, Syracuse University, Syracuse, New York 13244-1130 USA}

\author{Kartikeya Arora}
\affiliation{Department of Physics, Indian Institute of Technology (Banaras Hindu University), Varanasi 221005, India}
\affiliation{Département de Physique and Institut Quantique, Université de Sherbrooke, Sherbrooke J1K 2R1 QC, Canada}

\author{Peter L. McMahon}
\affiliation{School of Applied and Engineering Physics, Cornell University, Ithaca, NY, 14853, USA}

\author{B. L. T. Plourde}
\affiliation{Department of Physics, Syracuse University, Syracuse, New York 13244-1130 USA}

\author{Baptiste Royer}
\email{baptiste.royer@usherbrooke.ca}
\affiliation{Département de Physique and Institut Quantique, Université de Sherbrooke, Sherbrooke J1K 2R1 QC, Canada}

\author{Valla Fatemi}
\email{vf82@cornell.edu}
\affiliation{School of Applied and Engineering Physics, Cornell University, Ithaca, NY, 14853, USA}

\date{\today}

\begin{abstract} 

Spins and oscillators are foundational to much of physics and applied sciences.
For quantum information, a spin 1/2 exemplifies the most basic unit, a qubit. 
High angular momentum spins (HAMSs) and harmonic oscillators provide multi-level manifolds 
which have the potential for hardware-efficient protected encodings of quantum information and simulation of many-body quantum systems.
In this work, we demonstrate a new quantum control protocol that conceptually merges these disparate hardware platforms. 
Namely we show how to modify a harmonic oscillator on-demand  to implement a continuous range of generators to accomplish linear and nonlinear HAMS dynamics.
The spin-like dynamics are verified by demonstration of linear spin coherent (SU(2)) rotations, nonlinear spin 
rotations, and comparison to other manifolds like simply-truncated oscillators. 
Our scheme allows the 
universal control of a spin cat logical qubit encoding with interpretable drive pulses:
we use linear operations to accomplish four logical gates, and further show that nonlinear 
spin rotations can complete the logical gate set. 
Our results show how motion on a closed Hilbert space can be useful for quantum information processing and opens the door to superconducting circuit simulations of higher angular momentum quantum magnetism. 

\end{abstract}

\maketitle

Drive-induced Hamiltonian engineering is a powerful approach for designing nontrivial quantum dynamics.
This is particularly the case for harmonic modes that host nontrivial continuous variable quantum states but require an activated nonlinearity to create and manipulate those states~\cite{ma_quantum_2021, gross_hardware-efficient_2021}.
A popular approach to accomplishing this is to couple an ancilla qubit to a harmonic oscillator, where the implicit nonlinearity of the qubit enables activated nonlinear operations on the oscillator.
This oscillator-qubit approach has been gainfully used in particular in superconducting circuits (electrodynamic qubit and electrodynamic oscillator) ~\cite{albert_bosonic_2022, campagne-ibarcq_quantum_2020} and trapped ion systems (spin and phonon) ~\cite{fluhmann_encoding_2019,  de_neeve_error_2022, PhysRevLett.131.033604} to demonstrate non-classical bosonic control, including achievements of beyond-break-even error correction ~\cite{sivak_real-time_2023, ni_beating_2023}. 

In the context of superconducting circuits,  driving of a Josephson element has been ubiquitous in engineering operations that facilitate bosonic quantum error correction~\cite{leghtas_confining_2015,grimm_stabilization_2020, gertler_protecting_2021, ma_error-transparent_2020}. 
This approach provides flexible engineering of a variety of useful interaction Hamiltonians, but often does not provide universal control of the bosonic mode necessary for performing arbitrary logical gates 
without concatenating pulses with different parameters (i.e., requires a high circuit depth) or full numerical optimization of a time-domain pulse.
Existing approaches to universal oscillator control exploit the fact that in the dispersive regime, the qubit exhibits a manifold of oscillator-dependent transition frequencies~\cite{bretheau_quantum_2015,krastanov_universal_2015,heeres_cavity_2015,heeres_implementing_2017}. 
However, this picture is accurate only at low oscillator energies, which is in tension with canonical oscillator encodings that assume weight at infinite energy, such as so-called cat codes \cite{Mirrahimi_2014} and Gottesman-Kitaev-Preskill (GKP) grid codes~\cite{gottesman_encoding_2001}. 
The energy extent of these bosonic codes can also impose intrinsic limits on the fidelity of logical operations~\cite{hastrup_unsuitability_2021, 
2305.05262}. 
As alternatives, bosonic encodings such as the binomial codes~\cite{michael_new_2016} are designed to have a strictly finite extent in the oscillator Hilbert space. 
However, universal control in these contexts is usually accomplished by purely numerical or empirical optimization, resulting in limited intuition about the accomplished evolution~\cite{hofheinz_synthesizing_2009,heeres_cavity_2015,krastanov_universal_2015,heeres_implementing_2017,fluhmann_encoding_2019,eickbusch_fast_2022,reinhold_error-corrected_2020}.

\begin{figure*}
\centering
\includegraphics[width = 0.85\textwidth]{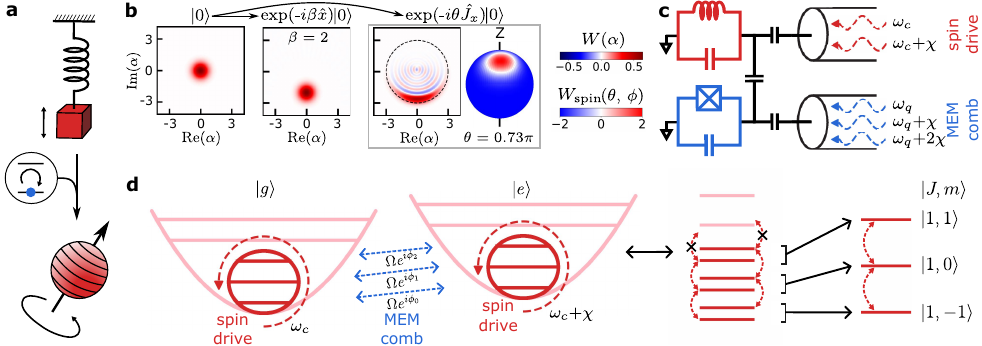}
\caption{ 
\textbf{Matrix element modification (MEM) of an oscillator.} 
(a) Cartoon illustrating our approach that takes a harmonic oscillator and transforms it into a spin using a scheme that incorporates an auxiliary qubit.
(b) Normally, a resonantly driven oscillator is displaced, e.g., by amplitude $\beta$ (=2, here), from the vacuum to a bosonic coherent state, as depicted by the first operation in the bosonic Wigner function representation $W(\alpha)$ with $\alpha$ being range of displacements. 
In this work, we instead aim to cause resonant drives to act with the generator of a spin, such as $\hat{J}_x$, which induces rotations by angles $\theta$ (=0.73$\pi$, here) into spin coherent states. 
This is shown via the second operation for the case of spin 9/2, represented by both the bosonic Wigner function $W(\alpha)$ and the spin Wigner function $W_{\mathrm{spin}}(\alpha)$~\cite{radcliffe_properties_1971}. 
The black dotted circle corresponds to a maximum radius for spin states $|\alpha| = \sqrt{N +1}$ where $N$ is the highest needed photon number.
(c) Our experimental setup consists of a qubit (blue; in this case a transmon) coupled to an oscillator (red).
The qubit is driven with a frequency comb to accomplish matrix element modification (the ``MEM comb") and cavity is driven to accomplish the spin drive.
(d) The MEM comb Rabi-splits the targeted cavity states and thereby blockades the cavity from exiting those states.
The key feature is the phases on each tooth of the MEM comb drive modify the matrix elements of the oscillator to that of a spin.
}
\label{fig:fig1} 
\end{figure*}

Here, we present a new approach for adiabatic universal oscillator control over a subspace of the oscillator, whereby the matrix elements of an isolated manifold of oscillator states are adjusted by design.
This allows for activated implementation of 
linear and nonlinear spin rotations through resonant driving, which provide an interpretable approach to universal control over the isolated manifold.
With this, we show how we can modify the oscillator to host SU(2) rotations (linear spin rotations) and thus transform it into a kind of high angular momentum spin (HAMS). 
We also show that nonlinear spin rotations, that have generators which are higher powers of linear spin rotation generators, can be accomplished in the oscillator.

We apply our method to accomplish logical gates on a spin cat encoding similar to those proposed in the context of nuclear spins~\cite{gupta_robust_2024, gross_hardware-efficient_2021}: 
the SU(2) rotations provide logical Pauli gates,
and we also identify nonlinear spin rotations that provide Hadamard and non-Clifford gates.
We demonstrate this experimentally on a circuit quantum electrodynamics-based platform, which is readily extensible, providing prospects for HAMS lattices to study frustrated magnetic systems and surface encodings. 

\section{Theoretical setup}

Our approach to synthetic HAMS dynamics requires changing the nature of the oscillator. 
Typically, a linear drive displaces the oscillator from the vacuum state(Fig.~\ref{fig:fig1}(b), left) to a coherent state of finite amplitude  (Fig.~\ref{fig:fig1}(b), middle). 
In contrast, for a spin, a linear rotation induces periodic dynamics of spin coherent states  (Fig.~\ref{fig:fig1}(b), right). 
Our goal, therefore, is to change the generator of unitary evolution from the unbounded, linear position or momentum operators ($\hat{x}$ or $\hat{p}$) to a finite-dimensional spin operator ($\hat{J_x}$ or $\hat{J_y}$).

To transform the oscillator to a HAMS, we first couple the oscillator to an ancilla qubit to develop a strong dispersive coupling strength $\chi$:
\begin{equation}
\begin{aligned}
    H / \hbar =  \omega_c \hat{a}^\dagger \hat{a} + \omega_q \ket{e}\bra{e} + 
    \chi \hat{a}^\dagger \hat{a} \ket{e}\bra{e}~,
\end{aligned}
\end{equation}
where $\omega_c$ is the cavity frequency, $\hat{a}$ is the cavity lowering operator, $\omega_q$ is the ancilla qubit frequency, and $\ket{e}$ is the ancilla qubit excited state (and $\ket{g}$ the ground state). 

Next, we apply a comb of Fock-state selective drives to the qubit at $\omega_q + n\chi$ for $n={0,1,...,N}$, as depicted in Fig.~\ref{fig:fig1}(c-d). 
The qubit drive Hamiltonian is 

\begin{equation}
\begin{aligned}
    H_{d_1}={} \Omega \sum_{n=0}^{N} \cos\left(\omega_q t +n \chi t + \phi_n + \frac{\pi}{2}\right)\hat{\sigma}_y 
    \label{eqn:qubit_drive}
\end{aligned}
\end{equation}
where $\Omega$ is the qubit Rabi drive rate, $t$ is the duration of the drive, and $\phi_n$ is the phase of the drive at $n\chi$ shifted frequency.
The number $N$ in such drive dictates the size of the isolated manifold of Fock states, which exists in two copies commensurate with the two levels of the qubit. 
Manifolds of similar size were isolated and driven in the past~\cite{bretheau_quantum_2015,signoles_confined_2014}, but in those implementations the size of the system also fixes the resulting dynamics.
For generating the dynamics in the oscillator we also apply a double-frequency drive to it,

\begin{equation}
    \begin{aligned}
        H_{d_2} = -i \epsilon \sum_{n=0}^1 \cos\left(\omega_c t + n \chi t + \varphi+ \frac{\pi}{2}\right) (\ha - \had)
    \end{aligned}
    \label{eqn:cavity_drive}
\end{equation}

 where $\epsilon$ is the cavity drive rate, and $\varphi$ is the common drive phase for cavity drive at both the frequencies. 
This double frequency drive accomplishes an unselective displacement of the cavity. 
The phase
of these drives, $\varphi$, is set to zero unless mentioned otherwise.

At the heart of our approach is the crucial distinction of tunability of the Rabi drive phases $\phi_n$, which allows us to tune the matrix elements within a given manifold to design its dynamics under drive.

This works because each oscillator Fock state $\ket{n}$ becomes tied to a qubit state rotating about the axis defined by $\phi_n$, e.g., $\ket{n,g} \rightarrow \ket{n}\otimes (\exp(-i\phi_n/2)\ket{g} + \exp(i\phi_n/2)\ket{e})/\sqrt2$.
After adiabatically eliminating the qubit dynamics in the limit $\epsilon\ll\Omega\ll|\chi|$, we can interpret the angles $\phi_n$ as modifying the expectation values of the raising and lowering operators:
\begin{equation}
\begin{aligned}
    \bra{n-1}\hat a \ket{n} \rightarrow \sqrt{n}\cos\left(\frac{\delta \phi_n}{2}\right)~,
\end{aligned}
\end{equation}
where we have defined $\delta \phi_n = \phi_n - \phi_{n-1}$.

This effect makes possible a continuum of generators $\hat{M} _\varphi$ of unitary rotations $\hat{U}=\exp\left(-i \theta \hat{M}_\varphi \right)$ for a manifold of $N+1$ levels in the oscillator:

\begin{equation}
\begin{aligned}
    \hat{M} &= \sum_{n=1}^{N}\sqrt{n} \cos\left(\frac{\delta \phi_n}{2}\right)\ket{n-1}\bra{n}, \\
    \hat{M}_\varphi &=  e^{-i\varphi} \hat{M} + e^{i\varphi} \hat{M}^{\dagger}~. \label{eq:Mphi}
\end{aligned}
\end{equation}

Finally, the rotation angle is given by $\theta = \epsilon t$.

\begin{figure*}[t]
\centering
\includegraphics[width = \textwidth]{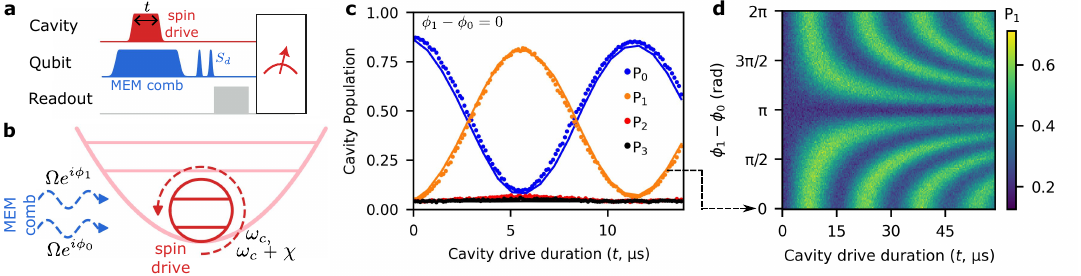}
\caption{\textbf{Minimal implementation: matrix-element-tunable spin 1/2.} (a) Pulse schematic for creating spin 1/2 dynamics in the cavity. MEM comb drive on qubit modifies the harmonic oscillator Hamiltonian to that of a spin and spin drive induces the spin rotations. $S_d$ is the ``decoder" SNAP gate used to disentangle the qubit and oscillator (see Appendix \ref{app:thy} for details). 
(b) Schematic of the experiment and drives: two Rabi drives at $\omega_q$, $\omega_q+\chi$ with amplitudes $\Omega e^{i\phi_0}, \Omega e^{i\phi_1}$, and two cavity drives at $\omega_c$, $\omega_c+\chi$ with equal amplitudes $\epsilon$. 
(c) Population of the cavity states for varying duration of cavity drives with $\epsilon/2\pi= \SI{94}{\kilo\hertz}$, $\Omega/2\pi= \SI{0.99}{\mega\hertz}$, $\chi/2\pi = \SI{-2.54}{\mega\hertz}$. Barring some small leakage, the Hilbert space is confined to the first two Fock states. 
(d) Oscillations of the population of $|1\rangle$ as a function of the qubit Rabi drive phase difference $\phi_1 - \phi_0$. 
At phase difference $\pi$, oscillations are stopped because the matrix element is zero. 
The cavity drive amplitude is constant across the entire plot and corresponds to $\epsilon/2\pi = \SI{72}{\kilo\hertz}$. For (d), the qubit drive duration is longer than (c), resulting in more decoherence and therefore less contrast in the oscillation.
}
\label{fig:fig2}
\end{figure*}

Commensurate with the construction, we consider the case of HAMS that have equally spaced energy levels. 
Harmonicity is useful for multilevel systems because it renders decayed photons indistinguishable regardless of the state of the system, which avoids dephasing for logical encodings.
For a spin with total angular momentum $J$, we set $N = 2J$ and then modify the generators of rotations to match that of a linear HAMS $\hat{M}_\varphi = \hat{J}_\varphi$, with a one-to-one mapping of Fock states to spin angular momentum states: $|n\rangle \leftrightarrow |J, m\rangle$, with $n = J + m$ , $m$ being the spin angular momentum along the $\hat z$ axis (see Appendix \ref{app:thy}). 
The following condition for the MEM comb phases accomplishes this:
\begin{equation}
    \phi_n = \phi_{n-1} + 2\cos^{-1}\left(\sqrt{\frac{2J + 1 - n}{2J}}\right)~. \label{eq2}
\end{equation}
Such dynamics accomplish SU(2) rotations.
Instead if all the qubit drive phases ($\phi_n$) are set to zero, then we observe dynamics equivalent to the blockaded oscillator dynamics demonstrated in Ref.~\cite{bretheau_quantum_2015}. We show a comparison of blockaded oscillator dynamics and spin dynamics in Appendix \ref{app:expt_comparisons}.

These SU(2) rotations have $\hat{J}_x$, $\hat{J}_y$, $\hat{J}_z$, $\mathbb{I}$ as their generators, so we call them linear spin rotations.
We can also accomplish nonlinear HAMS dynamics by choosing any phases that deviate from that of Eq.~\eqref{eq2}.
This is because generators of the form Eq.~\eqref{eq:Mphi} can be viewed as a linear combination of the generators
\begin{align}
    \hat{M} = \sum_{k=1}^{2J} c_k \left[ \hat{J}_-, J_z^k \right] ~, 
    \label{eq:M_nonlin}
\end{align}
where all $k>1$ are nonlinear terms. 
We refer to these rotations as nonlinear spin rotations as their generators include higher powers of the linear spin rotation generators.

Combining the SU(2) rotations with  nonlinear spin rotations provides universal control over the isolated manifold~\cite{merkel_quantum_2009}.
This can be expanded or concatenated to cover the oscillator Hilbert space (see Appendix \ref{app:universality}).
We note that this is an unusual object:  HAMSs in solid state systems (or, emulated HAMS) typically have unequal level spacing and in practice require separate drives addressing each spin-flip transition individually to accomplish universal control (see, e.g., Refs.~\cite{fernandez_de_fuentes_navigating_2024, gupta_robust_2024, doi:10.1126/science.1173440, Nguyen2024, PhysRevLett.125.170502}). 
In principle, our system remains with equal energy level spacing at all times,  and can accomplish universal control over an oscillator subspace purely through adjusting the matrix elements. 
Finally, we note that  real experimental hardware will have non-idealities, like violation of the adiabatic condition and inherited anharmonicity in the cavity, which we address in the discussion section. 
Next, we will describe our experimental implementations.

\section{Matrix-element tunable spin 1/2}

In our experiments we have used an aluminum $\lambda/4$ cavity as a harmonic oscillator and a transmon qubit as the ancilla.
Details of the package and device parameters are given in Appendix \ref{app:technical} (also, see Appendix \ref{app:methods}).

The simplest form of the experiment is to create a spin 1/2 degree of freedom. 
For this, we drive the qubit at the $n=0$ and $n=1$ Fock state shifted frequencies $\omega_q$ and $\omega_q + \chi$ ($\left(\ket{g} \leftrightarrow \ket{e}\right)\otimes\ket{n}$) to isolate the $n=0$, $1$ states of the cavity, as shown in Fig.~\ref{fig:fig2}(a-b). 
The cavity dynamics in these conditions and associated simulations are given in Fig.~\ref{fig:fig2}(c), verifying the isolation of a two level system, as was previously accomplished through a more conventional photon blockade~\cite{bretheau_quantum_2015}.

Unlike typical spins (or other qubits), here we have the freedom to tune the transition matrix element while keeping the transition frequency fixed. 
This is accomplished by tuning the relative phase of the two drives on the ancilla qubit, $\phi_1 - \phi_0$. 
By modifying the relative phase, the overlap of the effective eigenstates can be tuned, resulting in varying the Rabi rates as shown in Fig.~\ref{fig:fig2}(d), which follows the expectation from Eqs.~\eqref{eq:Mphi}-\eqref{eq2}.
The most extreme case, with the drives fully out-of-phase, $\phi_1 - \phi_0 = \pi$, suppresses rotations of the effective spin 1/2 despite the presence of the resonant drive.
Thereby, we accomplish the notion of matrix element modification.

\section{HAMS rotations}

\subsection{Linear spin rotations}

\begin{figure}
\centering
\includegraphics[width = \columnwidth]{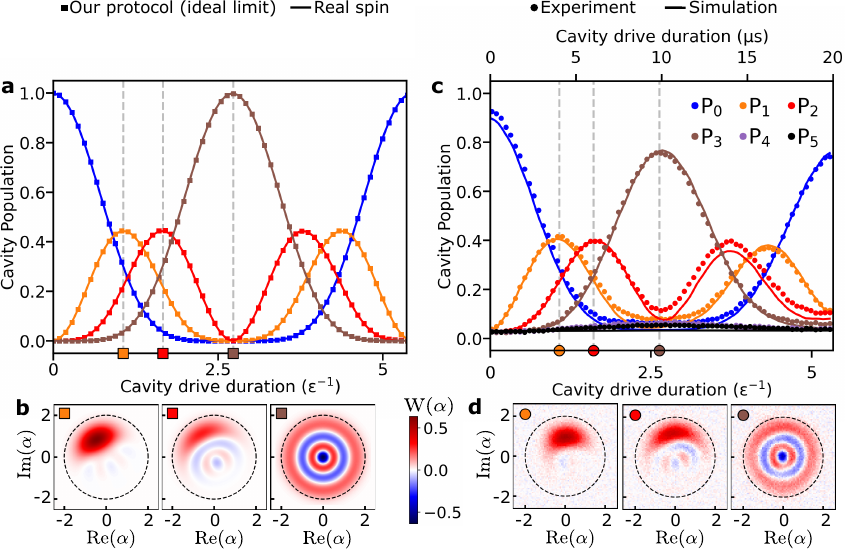}
\caption{\textbf{HAMS dynamics for a spin 3/2:}
(a) Dynamics of spin states of a driven spin 3/2 system. 
The points are numerical simulations of our protocol in the ideal limit $|\chi| \gg \Omega \gg \epsilon$. 
The solid line is for an ideal driven spin 3/2, which matches with our simulated spin. 
(b) The Wigner functions of the cavity state at the marked points for our spin protocol are shown.
(c) Experimentally measured probabilities of different spin states of a driven synthetic spin 3/2 in our cavity with $\epsilon/2\pi= \SI{80}{\kilo\hertz}$, $\Omega/2\pi=\SI{732}{\kilo\hertz}$, $\chi/2\pi= \SI{-3.56}{\mega\hertz}$. See Table.~\ref{tab:params_table}, column four, titled ``Measured values (2)", for details of the device parameters.
The experimental data is shown as dots, and the solid lines are numerical simulation of the spin protocol that considers experimental non-idealities and decoherences. 
The numerical simulation uses independently measured quantities for all inputs excepting subtle (about 3\%) differences in Rabi amplitudes.
(d) Experimentally measured Wigner functions at the marked points are shown. 
}
\label{fig:fig3}
\end{figure}

A synthetic HAMS with total spin $J$ requires $2J+1$ Fock states of the cavity to be isolated. 
In Fig.~\ref{fig:fig3} we show the spin dynamics of $J=3/2$ in both theory (panels a-b) and experiment (panels c-d). 
$J=1$ and $J=2$ are shown in Appendix \ref{app:expt_comparisons}.
For the theory, we show both the real spin dynamics (solid lines) and our protocol (markers) under ideal limits of $\epsilon  = 0.01\Omega = 10^{-4}|\chi|$, showing that the basic protocol functions as desired. 
At half period, the now-inverted spin coherent state is equivalent to a Fock state ($|3\rangle$), as shown in Fig.~\ref{fig:fig3}(b). 

The experimental implementation (Fig.~\ref{fig:fig3}(c-d)) exhibits the expected spin dynamics with the addition of energy losses and coherent infidelities (see Appendix for details), which are well captured by numerical simulation (lines). 
We also perform Wigner tomography by measuring Fock state parity after displacements~\cite{bertet_direct_2002}, and find comparable states to the ideal simulations. 
In particular, the Wigner at half period nearly exhibits the rotational symmetry in phase space that we expect. 
This is contrasted with conventional photon blockade~\cite{bretheau_quantum_2015} which exhibits a distinct and aperiodic quantum evolution (see Appendix~\ref{app:expt_comparisons}).

\subsection{Nonlinear spin rotations } \label{sec:nonlin}

Linear spin rotations are insufficient for universal computation for a Hilbert space of dimension larger than 2, including any logical qubit encoding in a multi-level system. 
Therefore nonlinear spin rotations, which we define as arising from generators that do not correspond to a linear combination of the SU(2) generators $J_x$, $J_y$, and $J_z$, are necessary to accomplish universal logical operations. 
Per Eq.~\eqref{eq:Mphi},~\eqref{eq:M_nonlin} we can accomplish nonlinear rotations that involve generators that are higher powers of the linear spin rotation generators. 
These rotations enable us to span the full SU(d) of the Hilbert space through concatenated operations~\cite{merkel_quantum_2009}.  
We promote here the possibility of using these nonlinear spin rotations (with generators described by Eq.~\eqref{eq:Mphi}, ~\eqref{eq:M_nonlin})
to accomplish novel logical control of a spin cat encoding (see section \ref{sec:logical_gate}).

Logical qubit encodings often require a specific parity structure in Fock space, e.g., the logical codewords for some binomial codes \cite{Hu2019,PhysRevX.6.031006} and cat codes \cite{Mirrahimi_2014, ofek_extending_2016} are even parit.y states. 
Therefore, any designed unitary operation on the logical codewords must respect this structure of parity.  
Here, we identify a class of nonlinear spin rotation that are periodic and preserve parity at $2\pi$ rotation while accomplishing a non-trivial operation.
For linear spin rotations $\hat{J}_x$ or $\hat{J}_y$, a $2\pi$ rotation results in the application of $\pm \mathbb{I}$, where the sign is $+$ for integer spin and $-$ for half-integer spin.
For half-integer spins we have identified a class of nonlinear spin rotation that preserves parity after a $2\pi$ rotation, where we use a unifying convention that $4\pi$ rotations produce the identity, just as in linear spin rotations. 
This class of nonlinear spin rotations can be used for accomplishing a continuous range of unitaries that can be applied for spin cat logical encoding of integer spin.

\begin{figure}
\centering
\includegraphics[width = \columnwidth]{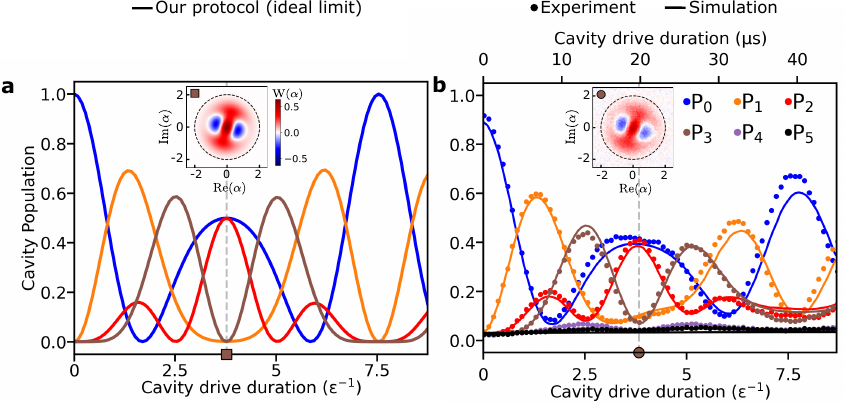}
\caption{\textbf{nonlinear spin dynamics that preserve parity at $2\pi$ rotation:}
(a) This is a numerical simulation of a lossless system in the ideal limit $|\chi| \gg \Omega \gg \epsilon$. 
The Wigner function of the cavity state at $2\pi$ rotation is shown in the inset.
(b) Experimental implementation of the nonlinear spin rotation, with measured probabilities for the nonlinear rotation shown in the figure. 
At $2\pi$ rotation, a spin-1 cat state is prepared from the ground state: $|0\rangle  \rightarrow (|0\rangle + |2\rangle)/\sqrt{2}$. 
The superposition phase can be adjusted by the phase of the cavity drive. 
The inset shows the experimentally measured Wigner function for this state. 
}
\label{fig:fig_gen_rots}
\end{figure}

As a demonstration, we consider an operation that accomplishes $\left(\ket{0} + \ket{2}\right) \bra{0} /\sqrt{2}  $. 
For this, we choose a $4\times4$ generator matrix that corresponds to a spin 3/2 system. 
Using the method of construction detailed in Appendix \ref{app:parity_preserving}, we can set the ratio of eigenvalues of this generator to be integer multiples of the fundamental eigenvalue to ensure a periodic unitary evolution. 
Then we can search the parameter space to find the required matrix elements for the desired unitary operation.  
In terms of the spin operators of Eq.~\eqref{eq:M_nonlin}, we obtain $(c_1, c_2, c_3) = (0.205, -0.034, 0.064)$.
Then, by modifying the phases $\phi_n$ according to Eq.~\eqref{eq:Mphi}, we obtain the generator for nonlinear spin rotation required to perform the said operation.
In Fig. \ref{fig:fig_gen_rots} we show the idealized and experimental evolution of the system under such a drive, indeed finding that it works as expected. 
In fact, there exist a continuous range of generators for such parity-subspace rotations that accomplish the operation
$\left(\cos(\gamma)\ket{0} +  \sin(\gamma)\ket{2}\right) \bra{0}+  \left(\sin(\gamma)\ket{0} -  \cos(\gamma)\ket{2}\right) \bra{2}  $ for $0\geq\gamma\geq\pi/2$.
For larger spins, this concept extends to rotations of the form $\left(\cos(\gamma)\ket{0} + \sin(\gamma)\ket{2N}\right) \bra{0} +
\left(\sin(\gamma)\ket{0} - \cos(\gamma)\ket{2N}\right) \bra{2} $ by using generators of size $(2N+2)\times(2N+2)$ where $N \in \mathbb{Z}_+$. 
We can also vary the cavity drive phase $\varphi$ to change the relative phase of the superposition of $\ket{0}$ and $\ket{2N}$ created by the nonlinear spin rotation.
Such rotations, and their generalizations, can be useful for achieving universal control of spin cat logical encodings which we discuss in the next section.

\section{Logical gates on spin cats}
\label{sec:logical_gate}
\begin{figure}
\centering
\includegraphics[width = \columnwidth]{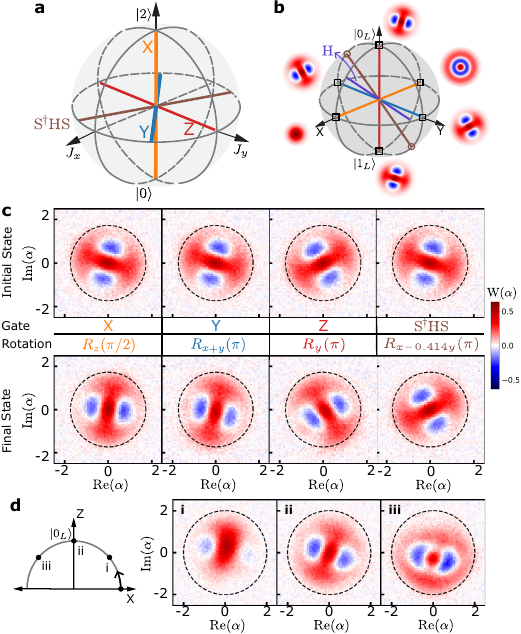}
\caption{\textbf{Logical operations on a spin kitten:}  
(a) A representation of the manifold of spin 1 coherent states, showing the axes around which we rotate the state for different spin kitten gates.
(b) The logical Bloch sphere for the two-legged spin kitten and the Wigner function of the states at the cardinal points. In purple color we also show the axis of rotation for a H gate that can be accomplished by a nonlinear spin rotation.
(c) The table shows the experimentally measured Wigner functions of the prepared spin kitten states (first row), the gates we apply on it (second row), the rotation that accomplishes this gate (third row) and the experimentally measured Wigner function of the final state we get after the gate is applied (fourth row).
For X, Y and $\textrm{SHS}^\dagger$ gate we start from a $|1_L \rangle$ spin kitten while for the Z gate we start from a $|0_L \rangle + i|1_L \rangle$ superposition state. 
Here we have defined $|0_L\rangle = (|0\rangle + |2\rangle)/\sqrt{2}$ and $|1_L\rangle = (|0\rangle - |2\rangle)/\sqrt{2}$ and ignored any global phases for the gates applied. 
The notation for rotation axis $R_{a+b}(\theta)$ should be read as rotation by angle $\theta$ around the axis $\hat{J_{a}} + \hat{J_{b}}$.
(d) We show three operations that change the magnitude of the superpositions between the logical states of the spin cat encoding using nonlinear spin rotations. 
On the left diagram, we show the location of the states created on a great circle on the XZ Plane of the logical Bloch sphere when starting from the marked $|+X\rangle$ state. 
Experimentally measured Wigner function for those states are shown in the following panels. 
The Wigner functions show  that different superpositions of logical states are created using nonlinear rotations.
The Wigner function in (ii) shows the action of a H gate on $|+X\rangle$ state indicating that nonlinear spin rotations can be used to achieve H gate.
}
\label{fig:fig4}
\end{figure}

Spin rotations add the potential for logical gate operations on logical qubit encodings in multi-level systems \cite{10.3389/fphy.2020.589504}. 
We demonstrate this on the polar two-legged spin kitten encoding hosted by a synthetic spin 1, equivalent to a minimal binomial encoding \cite{Hu2019} (for higher spins, we would call these polar two-legged spin cat encodings). 
The logical qubit states $|0_L\rangle$ and $|1_L\rangle$ are given by $|n=0\rangle \pm |n=2\rangle$ $(|J=1, m = -1\rangle \pm |J=1, m = +1\rangle)$ (see Fig.~\ref{fig:fig4}(b)). 
This code can, in principle, detect a single photon loss event; interestingly, an SU(2) rotation then accomplishes erasure recovery to the codespace (see Appendix~\ref{app:spincat}).
Note that such erasure detection and recovery is not possible for two-legged bosonic cats; this shows a qualitative distinction between two-legged spin cats and two-legged bosonic cats. 
In fact, we initialized these states by first preparing the Fock state $|1\rangle$, followed by a $\pi/2$ spin 1 rotation to instantiate a state in the logical code space (see Fig.~\ref{fig:sf_spincat}). 

A representation of the manifold of spin coherent states is given in Fig.~\ref{fig:fig4}(a), and the corresponding logical Bloch sphere is shown in  Fig.~\ref{fig:fig4}(b). 
The figure additionally shows the SU(2) rotation axes that map to all three Pauli gates and a rotated Hadamard gate. 
Three of these rotation axes are on the equator, while the fourth is equivalent to a frame shift.
Fig.~\ref{fig:fig4}(c) shows visually, through Wigner distributions, the accomplishment of these four logical gates on the spin 1 spin kitten encoding.
In Appendix \ref{app:spin_cat_gate_all} we show the application of these rotations on all cardinal points of the Bloch sphere.

Finally, we show that it is possible to use nonlinear spin rotations to accomplish a Hadamard gate ($H$ gate) and other states that mix the magnitudes of the weights on $|0_L\rangle$ and $|1_L\rangle$.
Note that these are spin 3/2 nonlinear spin rotations that are used for logical operations on a spin 1 spin cat.
In Fig.~\ref{fig:fig4}(d) we show Wigner function for three such operations on $|0\rangle$ along with the locations of final states on the logical Bloch sphere. 
These rotations can be interpreted as nonlinear spin rotations as shown in eqn. \ref{eq:M_nonlin} with coefficients $c_K$ given as - $(c_1, c_2, c_3) =$ (i)$(0.112, -0.060, 0.097)$, (ii) $(0.205, -0.034, 0.064)$, (iii) $(0.237, -0.004, 0.053)$.
Noting that in fact we have access to the full set of rotations on the Z-X great circle of the logical Bloch sphere (and any other amplitude-mixing great circle by changing the phase of the cavity drive), this protocol can provide universal logical control over this spin cat. 
By adding SWAP and exponential SWAP operations~\cite{liu_hybrid_2024}, universal logical computation over a multi-mode setup would also be possible.  

The set of four SU(2) rotation axes provides the same four logical gates for any HAMS encoding of the type  $|0\rangle \pm |N\rangle \leftrightarrow |-J\rangle \pm |+J\rangle$.
The same type of nonlinear spin rotation also complete the gate set for this type of encoding.
A set of three SU(2) rotations also accomplishes Pauli gates on a considerable subset of the binomial code words, which we will discuss in a later work.
To our knowledge, such a direct mapping of linear displacements to logical gates is not available for finite-energy versions of infinite-energy bosonic encodings: the native displacement operation only approximates logical gates for GKP encodings~\cite{gottesman_encoding_2001,fluhmann_encoding_2019,campagne-ibarcq_quantum_2020}, while for stabilized cat encodings it provides only one type of logical rotation~\cite{Mirrahimi_2014,puri_engineering_2017,touzard_coherent_2018,grimm_stabilization_2020}.

\section{ Discussion}

Above, we described and demonstrated a matrix element modification protocol to design the quantum evolution of a resonantly-driven harmonic oscillator, and focused the demonstration on SU(2) rotations and one class of nonlinear spin rotation acting on spin coherent states and a spin kitten encoding.
Our protocol is interpretable in the sense that novel generators of unitary evolution associated to resonant driving can be written down explicitly, with a direct relationship between drive parameters and the unitary evolution.
These generators can then be designed to accomplish a desired unitary operation in a single shot.
We believe this is the first demonstrated universal oscillator control protocol (over a subspace of the oscillator) that combines such interpretability with simultaneous drives on the oscillator and ancilla.
Typically, brute-force numerical optimization of drive amplitudes at every time-step, starting from random seeds, is used to derive a desired operation~\cite{heeres_implementing_2017,zhang_engineering_2019,ni_beating_2023}. 
Without simultaneous driving, one may instead alternate different types of pulses on the oscillator and ancilla~\cite{heeres_cavity_2015,fosel_efficient_2020,kudra_robust_2022,eickbusch_fast_2022, diringer_conditional-not_2024}, which have the benefit of interpretability of the different types of pulses. 
However, the multi-operation sequences for approximating a given target operation lose that interpretability, again requiring abstract numerical optimization.

The interpretability of our protocol is possible in part because it functions in an adiabatic limit. 
However, adiabatic manipulations based on a dispersive interaction require long drive durations, resulting in errors due to decoherence. 
So, in experiments we work in a regime that is not strictly adiabatic.
As a consequence of not strictly obeying the adiabatic limit, we can see leakage outside of (N+1) Fock codespace (about 1-3\%) and an imperfect spin dynamics. In Appendix \ref{app:non-adiabaticity}, we discuss more about the effect of decoherences and non-adiabaticity.
Therefore, future effort will focus on numerical optimization with parameters from our interpretable (and designable) protocol as seed (i.e., an initial guess for parameters to optimize) for performing high fidelity unitaries with a circuit depth of one.

We are inspired in particular by speed-ups 
while starting from the adiabatic limit parameters as a seed ~\cite{kudra_robust_2022,landgraf2023fast}. 
These approaches achieved SNAP gates with duration near $2\pi/\chi$, which for $\chi\sim\si{\mega\hertz}$ would allow for operations in around 0.1\% of the decay time of state of the art oscillators~\cite{ganjam_surpassing_2024}.
We can also seek to correct for certain intrinsic non-idealities, like from cavity self-Kerr, through additional drives~\cite{zhang_drive-induced_2022}.
Targeted operations will include full logical gate sets and error recovery operations for binomial and spin cat encodings and other qudit encodings based on HAMS \cite{gross_designing_2021, gross_hardware-efficient_2021, PRXQuantum.5.020355}.

Finally, we highlight opportunities in the quantum simulation of HAMS lattices~\cite{auerbach_interacting_1994,savary_quantum_2016,zhou_quantum_2017,broholm_quantum_2020, taie_observation_2022, Rosenberg2024}. 
In particular, frustrated magnetic systems are most commonly considered for the spin 1/2 case, in part because it maps best to relevant solid state and qubit-array experimental contexts. 
Theoretically, however, distinct predictions have been made in comparing full- vs half-integer spin lattices and the quantum-to-classical crossover as HAMS get larger. 
HAMS rotations interleaved with inter-oscillator operations like beamsplitter and exponential SWAP operation can be used for trotterized simulation of such HAMS lattices \cite{liu_hybrid_2024}.

These cases are less straightforward to study in purely qubit-based systems, and nuclear spin-based systems require different chemical elements for different sized spins.
Furthermore, our scheme allows new questions, like asking how strongly interacting and frustrated spin lattice dynamics may change when the sites are initialized with highly non-classical states at each site. 
Finally, for a given sized HAMS, we have only explored one type of linear operation (SU(2) rotations), whereas Equation~\eqref{eq:Mphi} implies that a continuum of more general linear operations are available. 

\par
Note added: In the preparation of our manuscript, we became aware of two complementary works on distinct hardware platforms that also demonstrate HAMS SU(2) dynamics and Schrodinger spin cats \cite{2405.15857, 2405.15494}.
We remark that these works used a different kind of multi-frequency control which can also implement nonlinear spin rotations of the type described here.

\section*{Data and code availability}
All data generated and code used in this work are available at: \textcolor{black}{\href{10.5281/zenodo.14639780}{10.5281/zenodo.14639780}}.

\begin{acknowledgments}

We acknowledge very helpful discussions with Nicholas Frattini and comments on the manuscript by Manuel Muñoz Arias, Or Katz, and Shyam Shankar. 
We thank Vikas Buchemmavari and Ivan Deutsch for pointing us to the SU(d) universal control proof by Merkel~\cite{merkel_quantum_2009}. 
We also acknowledge the contributions of Haoran Lu in setting up the cryogenic and microwave setups used in this experiment as well as Ryan Byrne for designing, coordinating the machining of, and assembling the copper bracket. 
We gratefully acknowledge Nord Quantique for the fabrication of the $\lambda/4$ superconducting cavity and MIT Lincoln Laboratory for supplying the Josephson traveling-wave parametric amplifier used in our experiments.

V. F. is thankful for the support of the Aref and Manon Lahham Faculty Fellowship that contributed to this work.
This work was performed in part at the Cornell NanoScale Facility, a member of the National Nanotechnology Coordinated Infrastructure (NNCI), which is supported by the National Science Foundation (Grant NNCI-2025233).
The authors acknowledge the use of facilities and instrumentation supported by NSF through the Cornell University Materials Research Science and Engineering Center DMR-1719875. 
P. M. and A. S. gratefully acknowledge financial support from NTT Research and a DURIP award with AFOSR award number FA9550-22-1-0080.
C. W. acknowledges the support of the Grainger Fellowship from the University of Chicago.
B. R. acknowledges support from the National Sciences and Engineering Research Council of Canada (NSERC), the Canada First Research Excellence Fund, as well as the Fonds de Recherche du Québec, Nature et Technologie (FRQNT).
B. L. T. P., B. C., E. Y., and C. L. acknowledge support by the U.S. Government under Army Research Office Grant No. W911NF-18-1-0106.

\end{acknowledgments}

\section*{Author Contributions}
V. F. and B. R. conceived and supervised the project. 
C. W. designed the experimental microwave hardware with input from V. F. and B. R., and S. R. designed the 3d cavity package.
B. C. and L. Z. accomplished the on-chip fabrication (transmon and readout modes) with assistance from  S. R., C. L., E. Y., V. F., and B. L. T. P..
A. S. and S. R. set up set the microwave lines and calibrated the experimental device.
S. R. performed the experiments and analyzed the data with guidance and feedback from A. S., C. W., B. R., and V. F..
Design of the logical gates was done by O. W. with guidance from V. F. and B. R..
Simulations were accomplished by S. R. and K. A. with guidance from A. S., C. W., V. F., and B. R..
The manuscript was drafted by S. R., B. R., and V. F., with input from all co-authors. 

\section*{Competing Interests}

The authors declare no competing interests.

\clearpage
\onecolumngrid

\renewcommand{\thefigure}{S\arabic{figure}}
\renewcommand{\thetable}{S\arabic{table}}

\balancecolsandclearpage
\clearpage
\onecolumngrid

\balancecolsandclearpage

\appendix
\balancecolsandclearpage
\setcounter{figure}{0} 
\setcounter{figure}{0}
\renewcommand{\thefigure}{S\arabic{figure}}
\renewcommand{\theHfigure}{S\arabic{figure}}
\section{Methods}
\label{app:methods}

\subsection{Experimental Setup}

In our experiment, an aluminum $\lambda/4$ cavity ($\omega_c/2\pi \approx\SI{6.04}{\giga\hertz}$) is used as the harmonic oscillator, and a transmon qubit ($\omega_q/2\pi \approx\SI{5.56}{\giga\hertz}$) coupled to it serves as the ancilla, shown schematically in Fig.~\ref{fig:fig1}(b)~\cite{reagor_quantum_2016}.
Such a setup has been used in the past for demonstration of many continuous variable quantum information encodings and beyond-break-even error correction~\cite{albert_bosonic_2022,sivak_real-time_2023}.
The dispersive coupling $\chi/2\pi \approx \SI{-2.54}{\mega\hertz}$ between the transmon and the cavity provides the necessary setting to implement our protocol.
Details about our experimental setup, package, mode frequencies, coupling rates, and decoherence rates are given in Appendix~\ref{app:technical}, particularly Fig.~\ref{fig:sf_wiring} and Tab.~\ref{tab:params_table}.

\subsection{Details of Pulse Sequence}
\label{app:pulse_details}

The protocol starts with a multi-frequency drive on the qubit.
This drive has a $\sim$\SI{150}{\nano\second} rise and $\sim$\SI{150}{\nano\second} fall duration and a flat top portion in between (slightly different rise and fall duration for different spins). 
The rise and fall have a Gaussian shape. The Gaussian pulse stops and ends abruptly on either side of the center beyond 2.5 $\sigma$ of the pulse, where $\sigma$ is the standard deviation of the Gaussian distribution.

The flat top portion ensures that the Rabi rates are constant at all times while the cavity is being driven. While the flat-top qubit drive is on, we turn on the double-frequency cavity drive which also has a \SI{400}{\nano\second} Gaussian rise and fall and a flat top portion. 
The duration of this pulse is varied for different rotation of the HAMS as shown in Fig.~\ref{fig:fig2} such that the pulse always ends at the same time.
So, the pulse starts earlier if it is longer. 
This is done so that the duration between state creation and probing is same for all duration of cavity drive.
After turning off the qubit drive, we disentangle the qubit from the cavity states using a SNAP gate \cite{PhysRevLett.115.137002} consisting of two \SI{2}{\micro\second} long $\pi$ pulses to add appropriate phases to different cavity states.
The SNAP gate adds phases $\phi_n/2$ to the cavity states to accomplish this disentanglement where $\phi_n$ is chosen according to Eq.~\eqref{eq2}. 

Then we readout the qubit state and store it for post-processing. 
After that we probe the cavity population using number-selective $\pi$-pulses on the qubit and then measuring the state of the qubit. 
Finally we post-select on the cases where our first readout measured the qubit in its ground state. 
This is because we started the qubit in the ground state and after driving it for a few Rabi cycles it should end in the same state unless some decay or dephasing event has happened.
The cases where such events occur are ignored.
For experiments checking the spin evolution (e.g., Fig.~\ref{fig:fig3}), 15\% to 35\% shots are rejected from shorter times to longer times. 
For spin kitten gate experiments, about 10-15\% shots are rejected.

\subsection{Numerical Simulation}

Here we describe the numerical simulation method and parameters used for simulating the cavity populations observed in experiment. 
We work in a frame rotating at the cavity and qubit frequency and account for cavity-qubit dispersive coupling ($\chi$), second order dispersive coupling ($\chi^\prime$), and cavity self-Kerr ($K$), in the calculation. 
We treat the transmon qubit as an anharmonic oscillator and model 4 of its levels and include 2J+4 levels of the harmonic oscillator (cavity states) in the numerical calculation.
The annihilation operators of the transmon qubit and cavity are denoted by $\hat q$ and $\hat c$, respectively.
Under the multi-frequency cavity and qubit drives described in App. \ref{app:pulse_details}, the system Hamiltonian is given by 
\begin{equation}
    \begin{aligned} 
    \hH/\hbar = \chi \hat{n}_c \hat{n}_q + K \hat{c}^{\dag} \hat{c}^{\dag} \hat{c} \hat{c} /2 + \alpha \hat{q}^{\dag} \hat{q}^{\dag} \hat{q} \hat{q} /2 \\ 
    + \chi^{\prime} \hat{c}^{\dag} \hat{c}^{\dag} \hat{c} \hat{c} \hat{n}_q/2 \\
    +  \sum_{n=0}^{2J} \frac{\bar \Omega_k}{2} \expo{i( n \chi t+ \phi_n)} \hat q + \mathrm{h.c.} \\
    + \frac{\bar \epsilon}{2}[\expo{i\varphi} + \expo{i(\chi t + \varphi)}]\hat{c} + \mathrm{h.c.},
    \end{aligned}
\end{equation}
where $\alpha$ is the transmon qubit anharmonicity, $\bar \Omega_k$ is the qubit Rabi rate of the drive at frequency $\omega_q + n\chi$, $\bar \epsilon$ is the cavity drive rate and $\hat{n}_q = \hat q^\dag \hat q$, $\hat{n}_c = \hat c^\dag \hat c$.

We perform a numerical simulation using open source software package QuTiP~\cite{johansson_qutip_2013} and solve the Lindblad master equation 
\begin{equation}
    \begin{aligned}
    \dot{\rho}(t) ={}& -\frac{i}{\hbar} [H(t), \rho(t)] \\ &+ \sum_{n} \frac{1}{2}[2C_n\rho(t)C_n^{\dag} - \rho(t)C_n^{\dag}C_n
    -C_n^{\dag}C_n\rho(t)],
    \end{aligned}
\end{equation}
where $C_n = \sqrt{\gamma_n} A_n$. $\gamma_n$ are the loss rates and $A_n$ are the operators through which the system couples to environment and losses occur. We have used $\hat{c}$, $\hat{n}_c$, $\hat{q}$ and $\hat{n}_q$ as $A_n$ and $1/T_{1c}$, $1/T_{\phi c}$, $1/T_{1q}$, $1/T_{\phi q}$ as $\gamma_n$ respectively.
For the simulation we use the independently measured experimental parameters listed in Tab.~\ref{tab:params_table}.
As in experiments, we also choose a length of qubit Rabi drive longer than the maximum cavity drive and slightly vary it so that the qubit ends in $|g\rangle$ with high probability. 
Then we post-select on the cases where qubit ends in $|g\rangle$ at the end of the protocol. 
In simulations we include 6.5\% probability of readout misassignment and 5\% infidelity for the cavity Fock state selective $\pi$ pulse that is used for probing the population in the cavity.

For the simulations in Fig.~\ref{fig:fig3}, \ref{fig:fig_gen_rots},  \ref{fig:sf_spin1.5} and \ref{fig:sf_spin2} we have used slightly varying (2-3\%) $\Omega_k$ (={1, 1.03, 1.03, 0.97, 0.98}), which we found provided the best match, and which are consistent in scale with a 1\% amplitude variation of these drives seen on a spectrum analyzer. 

\section{Theoretical construction of the matrix element modification protocol} \label{app:thy}

\subsection{Matrix element modification protocol}

In the two-level approximation for the transmon qubit, the qubit-cavity system can be described by the Hamiltonian~\cite{blais_circuit_2021}

\begin{equation}
    \begin{aligned}
    \hH/\hbar ={}& \omega_q \ketbra{e} + \omega_c \hn_c + \chi \hn_c \ketbra{e} \\
&+  \Omega(t) \sy - i\epsilon(t)(\hat{c} - \hat{c}^{\dagger})\\
    \Omega(t) ={}& \bar \Omega \sum_{n=0}^{2J} \cos\left(\omega_q t +n \chi t + \phi_n + \frac{\pi}{2}\right)\\
    \epsilon(t) ={}& \bar \epsilon \sum_{n=0}^1 \cos\left(\omega_c t + n \chi t + \varphi+ \frac{\pi}{2}\right)~,
    \end{aligned}
\end{equation}

$\hat{n}_c = c^{\dagger}c$ is the number operator for the cavity (oscillator) with eigenvalues $n$ and eigenkets $|n\rangle$, and $\sigma_{x, y, z}$ are the Pauli operators.
The qubit and the cavity are both driven with frequency combs containing $2J+1$ and 2 components, respectively. For the qubit drive, the amplitude of each comb is set to be the same $\bar \Omega$, while the phase of each component $\phi_n$ is set to values that will yield the matrix element modification. For the cavity drive, we set the amplitude $\bar \epsilon$ and phase $\varphi$ of both frequency component the same.
In the joint rotating frame of both the cavity and the qubit, obtained after unitary transformation 

\begin{equation}
\hat U_1 = \exp[i \omega_q t \ketbra{e} + i\omega_ct \hn_c] 
\end{equation}

we get

\begin{equation}
    \begin{aligned}
    \hH_1 \approx{}& \chi \hn_c \ketbra{e}\\
    &+ \frac{\bar \Omega}{2} \sum_{n=0}^{2J} \expo{i( n \chi t+ \phi_n)} \sm + \mathrm{h.c.}\\
    &+ \frac{\bar \epsilon}{2}[\expo{i\varphi} + \expo{i(\chi t + \varphi)}]\hat{c} + \mathrm{h.c.},
    \end{aligned}
\end{equation}

$\sigma_{-} = \sigma_x - i\sigma_y$ and we neglected terms rotating at $\sim 2\omega_q,2\omega_c$ through a rotating wave approximation (RWA), which assumes that $\omega_q,\omega_c \gg |\chi|$. The Hamiltonian above is obtained through the transformation $\hH' = \hU \hH \hU^\dag - i \hU \dot{\hU}^\dag$.
Note that we also started with a two-level approximation for the transmon qubit, which implicitly assumes that the qubit drive is much smaller than the anharmonicity of the qubit, $\bar \Omega \ll \alpha$.

To gain insight into the dynamics under this driven Hamiltonian, we express the Hamiltonian in the interaction picture, \emph{i.e.} we transform the Hamiltonian according to the unitary 

\begin{equation}
\hat U_2 = \exp\left(i \chi t \hat n \ketbra{e}\right)    
\end{equation}

In the regime of interest where $\chi \gg \bar \Omega$, we perform another RWA and we obtain an approximate Hamiltonian
\begin{equation}
\begin{aligned}
    \hH_2 \approx{}& \frac{\bar \Omega}{2} \sum_{n=0}^{2J} \ketbra{n}(\sx \cos{\phi_n} - \sy \sin{\phi_n})\\
    &+ \frac{\bar \epsilon \expo{i\varphi}}{2}\sum_{n=1}^\infty \sqrt{n}\ket{n-1}\bra{n} + \mathrm{h.c.}.
\end{aligned}
\end{equation}
In this interaction frame and in the absence of a cavity drive ($\bar \epsilon = 0$), the eigenstates of the Hamiltonian for the first $2S$ Fock states will be of the form $\ket{n} \otimes\, \exp(-i \sz \phi_n/2 )\ket{\pm}$, where $\ket{\pm} = (\ket{g} \pm \ket{e})/\sqrt 2$. Going to a frame where all the qubit eigenstates are aligned along the $\sx$ axis, which we get through a unitary transformation 
\begin{equation}
   \hat U_3 = \exp\left(i \sum_{n=0}^{2J} \phi_n \ketbra{n}\otimes \sz /2\right), 
\end{equation}
we obtain 
\begin{equation}
\begin{aligned}
    \hH_3 \approx{}& - \frac{\bar \Omega}{2} \sum_{n=0}^{2J} \ketbra{n}\sx \\
    &+ \frac{\bar \epsilon \expo{i\varphi}}{2}\sum_{n=1}^{2J} \sqrt{n}\ket{n-1}\bra{n}\expo{-i \delta\phi_n \sz /2} + \mathrm{h.c.}\\
    & + \frac{\bar \epsilon \expo{i\varphi}}{2} \sqrt{2J+1}\ket{2J}\bra{2J+1} + \mathrm{h.c.}\\
    &+ \frac{\bar \epsilon \expo{i\varphi}}{2}\sum_{n=2J+2}^\infty \sqrt{n}\ket{n-1}\bra{n} + \mathrm{h.c.},
\end{aligned}
\end{equation}
where we have defined $\delta \phi_n \equiv \phi_n - \phi_{n-1}$.
Note that the third line is isolated from the last summation because it will disappear in the final approximation.

Finally, taking into account the qubit rotation due to the first term through a unitary transformation 
\begin{equation}
\hat U_4 = \exp(-i \bar \Omega t/2 \times \sum_{n=0}^{2J} \ketbra{n}\sx)
\end{equation}
and performing one last RWA assuming that $\bar \Omega \gg \bar \epsilon$, we obtain
\begin{equation}\label{eq:MatrixModHamiltonian}
\begin{aligned}
    \hH_f \approx{}& \frac{\bar \epsilon \expo{i\varphi}}{2}  \sum_{n=1}^{2J}\sqrt{n} \cos\left(\frac{\delta \phi_n}{2}\right)\ket{n-1}\bra{n} + \mathrm{h.c.}\\
    &+ \frac{\bar \epsilon \expo{i\varphi}}{2}\sum_{n=2J+2}^\infty\sqrt{n}\ket{n-1}\bra{n} + h.c. ~,
\end{aligned}
\end{equation}
where we obtain a first manifold ($n \leq 2J$) where transition matrix elements between neighboring Fock states can be tuned through a choice of $\delta\phi_n$, and a second manifold ($n \geq 2J+1$) where the matrix elements remain unchanged. 
Through a photon blockade process~\cite{bretheau_quantum_2015}, these manifolds stay separated, hence the absence of a matrix element between states $\ket{2J}$ and $\ket{2J+1}$. 

Note that after all the unitary transformations performed to obtain $\hH_f$, the evolution with modified matrix elements does not occur in the lab frame. As a result, expectation values should be computed by properly taking these transformations into account. Defining the total unitary transformation $\hU_{\mathrm{tot}} = \hU_4\hU_3\hU_2\hU_1$, where the time dependence has been kept implicit. In the lab frame, expectation value of an observable $\hat O$ and with an initial state $\rho$ is computed from
\begin{equation}
   \langle \hat O \rangle  = \mathrm{Tr}\left[\hat O \hU_{\mathrm{tot}}^\dag(t_f)\expo{-i t_f\hH_f}\hU_{\mathrm{tot}}(0)\rho\hU_{\mathrm{tot}}^\dag(0) \expo{i t_f\hH_f} \hU_{\mathrm{tot}}(t_f)\right]. 
\end{equation}
At $t=0$, the unitary transformation simplifies to $\hU_{\mathrm{tot}} = \hU_3$. Ideally, one would also choose a final time such that $\chi t_f = 2\pi k,\bar \Omega t_f = 2\pi l $ for $k,l \in \mathbb Z$. In that situation, and ignoring the effect of the first rotating frame transformation, we get
\begin{equation}
   \langle \hat O \rangle  = \mathrm{Tr}\left[\hat O \hU_3^\dag\expo{-i t_f\hH_f}\hU_3\rho\hU_3^\dag \expo{i t_f\hH_f} \hU_3\right]. 
\end{equation}
We finally obtain lab-frame evolution under the effective matrix-element modified Hamiltonian Eq.~\eqref{eq:MatrixModHamiltonian} by actively applying an entangling $\hU_3^\dag$ and disentangling ($\hU_3$) between the qubit and oscillator before and after the evolution, respectively. 
These operations can be realized by, for example, SNAP pulses~\cite{PhysRevLett.115.137002} which are designed to apply a unitary of the form of $\hU_3$. With these additional SNAP pulses applied before and after the evolution, we obtain in the ideal limit
\begin{equation}
   \langle \hat O \rangle  = \mathrm{Tr}\left[\hat O \expo{-i t_f\hH_f}\rho \expo{i t_f\hH_f}\right]. 
\end{equation}

In the experimental results presented in this manuscript, the condition on the final time to cancel the unitary transforms $\hU_2$ and $\hU_4$ is not perfectly met, and we have rather
\begin{equation}
   \langle \hat O \rangle  = \mathrm{Tr}\left[\hat O \hU_3\hU_{\mathrm{tot}}^\dag(t_f)\expo{-i t_f\hH_f}\rho \expo{i t_f\hH_f} \hU_{\mathrm{tot}}(t_f)\hU_3^\dag\right]. 
\end{equation}

Summarizing the assumptions made above in order for the various RWAs made above to work, we need the following parameter hierarchy:
\begin{equation}
    \omega_q, \omega_c \gg |\chi| \gg \bar \Omega \gg \bar \epsilon \gg \gamma_q,\kappa~,
\end{equation}
where $\gamma_q$, $\kappa$ are the loss rates of the qubit and cavity, respectively.

\subsection{Oscillator-to-HAMS mapping}
We first set our notation for the HAMS . We denote the angular momentum operators for a spin $J$ to be $\hat J_x^{(J)},\hat J_y^{(J)},\hat J_z^{(J)}$. For spin states we work in the basis of the $z$ angular momentum operator eigenstates, with $\hat J_z^{(J)}\ket{J,m} = m \ket{J,m}$. We also define the spin operators
\begin{equation}
    \hat J_-^{(J)} = \sum_{m=-J}^J \sqrt{J(J+1) - m(m-1)}\ket{m-1}\bra{m},
\end{equation}
with $\hat J_+^{(J)} = (\hat J_-^{(J)})^\dag$. The angular momentum operators can be expressed as $\hat J_x^{(J)} = (\hat J_-^{(J)} + \hat J_+^{(J)})/2$ and $\hat J_y^{(J)} = i(\hat J_-^{(J)} - \hat J_+^{(J)})/2$

We choose an oscillator-to-HAMS mapping where the vacuum state is mapped to the state $\ket{J,m = -J}$, and Fock states above are mapped to the other eigenstates of $\hat J_z^{(J)}$.

\begin{equation}
    \ket{n} \leftrightarrow \ket{J, m = n-J},
\end{equation}
which induces the mapping $\hn \leftrightarrow \hat J_z^{(J)} + J$. We denote the projector onto the $(2J+1)$-dimensional spin manifold
\begin{equation}
\hPi_J \equiv \sum_{n=0}^{2J} \ketbra{n}.
\end{equation}

Note that the mapping in this work is flipped with respect to the Holstein-Primakoff HAMS-oscillator mapping, which maps $\ket{n=0} \leftrightarrow \ket{J,m = +J}$. To obtain an effective Hamiltonian proportional to the angular momentum operator, we choose for $n \geq 1$ 
\begin{equation}
    \phi_n = \phi_{n-1} + 2\cos^{-1}\left(\sqrt{\frac{2J + 1 - n}{2J}}\right).
    \label{eqn:phase_modification}
\end{equation}
Since only the phases difference matter, we set $\phi_0 = 0$ for simplicity. Moreover, as shown in Eq.~\eqref{eq:MatrixModHamiltonian}, the sign of the phase differences has no impact (to first order), and we choose a positive phase difference. Under this choice of phase, Eq.~\eqref{eq:MatrixModHamiltonian} becomes
\begin{equation}
\begin{aligned}
    \hH_f \approx{}& \frac{\bar \epsilon \expo{i\varphi}}{2\sqrt{2J}}  \hat J_-^{(J)} + \mathrm{h.c.}\\
    &+ \sum_{n=2J+2}^\infty\epsilon(t)\sqrt{n}\ket{n-1}\bra{n} + h.c,
\end{aligned}
\end{equation}
Choosing the (global) cavity drive phase $\varphi$ allows to interpolate between $\hat J_x^{(J)}$ and $\hat J_y^{(J)}$. Choosing $\varphi = 0$ and projecting the Hamiltonian in the desired manifold, we obtain $\hat \Pi_J \hH_f \hat \Pi_J \propto \hat J_x^{(J)}$.

\subsection{Universality of matrix-modification scheme}
\label{app:universality}

\subsubsection{Givens Rotations}
While the matrix modification scheme presented here can only couple neighbouring oscillator levels, access to generators of the form
\begin{equation}
\begin{aligned}
    \hat{M} &= \sum_{n=1}^{2J}\sqrt{n} \cos\left(\frac{\delta \phi_n}{2}\right)\ket{n-1}\bra{n} \\
    \hat{M}_\varphi &=   e^{-i\varphi} \hat{M} + e^{i\varphi} \hat{M}^{\dagger} ~.
\end{aligned}
\end{equation}
is sufficient to obtain universal $\mathrm{SU}(N+1)$ control over the desired manifold (Fock states below $n \leq N = 2J$). For example, setting $\delta \phi_n = \pi$ leads to nullified matrix elements, such that choosing $\delta \phi_m = 0$ and $\delta \phi_{n \neq m} = \pi$ leads to full $\mathrm{SU}(2)$ control over the manifold spanned by the Fock states $\{\ket{m-1},\ket{m}\}$. In other words, we can perform Givens rotations between neighboring levels.
In particular, swap gates between neighboring levels can be engineered, which means that control of neighboring levels can be extended to full $\mathrm{SU}(2)$ control between any pair of Fock states. 
These pairwise unitaries, equivalent to the set of nearest-neighbor Givens rotations, can then be combined to obtain full $\mathrm{SU}(N+1)$ control~\cite{reck_experimental_1994}. 
We leave the optimal factorization of a general unitary into unitaries generated by $\hat M_\varphi$ operators for future work.

\subsubsection{SU($d$) control}

Our scheme can go beyond Givens-style rotations in that it can achieve universal control over any nearest set of $d$ levels of the system using just spin-$J$ SU(2) rotations and one more generalized generator $ \hat{M}_\varphi$~\cite{merkel_quantum_2009}. 
In particular, any $\hat{M}_{\varphi}$ which satisfies
\begin{equation}
    \begin{aligned}
        \mathrm{Tr}\left(\hat{M}_{\varphi}T_{q}^{(2)}(J)\right) \neq 0
    \end{aligned}
    \label{eqn:universalCond}
\end{equation}
 where $d = 2J + 1$, for some $q \in \{-2, -1, 0, 1, 2\}$, will generate $d$-level universal control when combined with the $\hat{J}_{x}^{(J)}$ and $\hat{J}_{y}^{(J)}$ operators we already demonstrated. 
 Here, $T_{q}^{(2)}(J)$ is any rank-2 irreducible spherical tensor, which are more generally given for any rank $k$ by
\begin{equation}
    \begin{aligned}
        T_{q}^{(k)}(J) = \sqrt{\frac{2k + 1}{2J + 1}}\sum_{m = -J}^{J}C_{k, q; J, m}^{J, m + q}|J, m + q\rangle\langle J, m|,
    \end{aligned}
    \label{eqn:sphericalTensors}
\end{equation}
where $C_{k, q; J, m}^{J, m + q} = \langle J, m + q | k, q; J, m\rangle$ is a Clebsh-Gordan coefficient for adding a spin-$J$ and spin-$k$ system to get a spin-$J$ sector. The two indices $q$ and $k$ are always integers satisfying $-k \leq q \leq k$ and $0 \leq k \leq 2J$.

We note that this condition for the extra $\hat{M}_{\varphi}$ generator is not stringent since nearly all of the $\hat{M}_{\varphi}$ generically satisfy Eq.~\eqref{eqn:universalCond}. 
Recall that Eq.~\eqref{eq:M_nonlin} intuitively shows that when $\hat{M}_\varphi$ deviates from SU(2) it must be considered a nonlinear operation. 
From the form of Eq.~\eqref{eqn:sphericalTensors}, we can see that any irreducible tensor $T_{q}^{(k)}(J)$ with $q = \pm1$ will only have matrix elements on the first above/below off-diagonal. Thus, for instance, simply setting $\delta\phi_{1} = 0$ and all the other $\delta\phi_{n} = \pi$ will always yield a $\hat{M}_{\varphi}$ that satisfies Eq.~\eqref{eqn:universalCond}.

This claim of $d$-level universality follows directly from a theorem by Merkel~\cite{merkel_quantum_2009}, which states that, for any $J$ and any $\hat{M}_{\varphi}$ satisfying Eq.~\eqref{eqn:universalCond}, the three operators $\{\hat{J}_{x}^{(J)}, \hat{J}_{y}^{(J)}, \hat{M}_{\varphi}\}$ generate the entire Lie algebra $\mathfrak{su}(d)$ under linear combinations and commutators, which is the standard necessary and sufficient condition for achieving full SU($d$) control under exponentiation.
We note how this presents a direct parallel with the case of bosonic control, in which linear displacements need only be supplemented by a single nonlinearity to accomplish universal control~\cite{liu_hybrid_2024}. 
The parallel is consistent with the existence of mappings between HAMS and bosons for the limit of $J\rightarrow\infty$.

\subsection{Parity-preserving operations via nonlinear spin rotations}
\label{app:parity_preserving}
In the main text, we focused on the nonlinear spin rotations that have periodic evolution and preserve parity at $2\pi$ rotation while accomplishing non-trivial operation on spin cat logical encodings.
Here we discuss the construction of such nonlinear spin rotations.
We consider even-dimensional generators $\hat{M}_\varphi$ with eigenvalues $\{\lambda\}$, which come in $\pm$ pairs.
We restrict to cases where the set of eigenvalues are integer multiples of the fundamental (smallest) eigenvalue $\{\lambda\} / |\lambda_\mathrm{min}| \in \mathbb{Z}$. 
The fundamental eigenvalue therefore sets the periodicity of unitary rotations $\hat{U} = \exp{\left(i\theta \hat{M}_\varphi\right)}$, such that $\hat{U}=\hat{I}$ when $\theta\lambda_\mathrm{min} = \pi$. 
When all eigenvalues are odd integer multiples of the fundamental, a $2\pi$ rotation (given by $\theta\lambda_\mathrm{min} = \pi$) results in $\hat{U}=-\hat{I}$ just as in the case of an SU(2) rotation for half-integer spins. 
If the eigenvalues are all integer multiple of the fundamental, such a $2\pi$ rotation results in more general operations that are also block diagonal in parity i.e. they preserve parity of the state after $2\pi$ rotation.
Such generators are necessarily nonlinear, since they never correspond to SU(2) rotations.
We can use Eq.~\eqref{eq:M_nonlin} to interpret this generator as a nonlinear spin rotation as the generators are a combination of higher powers of the linear spin rotation generators.

We give a formal proof that this class of nonlinear spin rotations used in the main text preserves parity of the state at $2\pi$ rotation.
 We have introduced parity-preserving operations, whereby we choose the eigenvalues of the generator $\hat{M}_{\varphi}$ to be integer multiples of the lowest eigenvalue, $\lambda / |\lambda_\mathrm{min}| \in \mathbb Z$. Since the generator $\hat{M}_{\varphi}$ induces transitions between neighboring Fock states, we have
\begin{equation}
    \hat P \hat{M}_{\varphi} \hat P = - \hat{M}_{\varphi}~,
\end{equation}
where we have defined the photon parity operator $\hat P = \mathrm{exp}(i \pi \hat a^\dag \hat a)$. This equation also implies that $\hat P \exp(i \theta \hat{M}_{\varphi}) \hat P  = \exp(- i \theta \hat{M}_{\varphi}) = [\exp(i \theta \hat{M}_{\varphi})]^\dag$. 
Choosing $\theta= 2\pi / |\lambda_\mathrm{min}|$, with $\lambda_\mathrm{min}$ being the eigenvalue with the lowest magnitude, we can express
\begin{equation}
    \begin{aligned}
        \hat P \mathrm{e}^{i \frac{2\pi}{|\lambda_\mathrm{min}|} \hat{M}_{\varphi}} \hat P &= \left[\mathrm{e}^{i \frac{2\pi}{|\lambda_\mathrm{min}|} \hat{M}_{\varphi}}\right]^\dag,\\
        &= \left[\sum_j \mathrm{e}^{i 2\pi \frac{\lambda_j}{|\lambda_\mathrm{min}|}} \ketbra{\lambda_j}\right]^\dag,\\
        &= \sum_j \mathrm{e}^{-i 2\pi \frac{\lambda_j}{|\lambda_\mathrm{min}|}}\ketbra{\lambda_j},
    \end{aligned}
\end{equation}
where we have labeled the eigenvalues and corresponding eigenkets by $\lambda_j$. In the special case where $\lambda_j/|\lambda_\mathrm{min}| \in \mathbb Z$, the acquired phases for all eigenkets are integer multiples of $\pi$, such that $\hat P \mathrm{e}^{i \frac{2\pi}{|\lambda_\mathrm{min}|} \hat{M}_{\varphi}} \hat P  =\mathrm{e}^{i \frac{2\pi}{|\lambda_\mathrm{min}|} \hat{M}_{\varphi}} $. In this situation, the parity operator commutes with the resulting unitary, which consequently preserves the parity of the photon number.

\balancecolsandclearpage

\section{Technical aspects of the experiment } \label{app:technical}

\subsection{Experimental hardware}
Schematic design of the package is shown in Fig.~\ref{fig:sf_freq}(b). 
In this section we describe the production and characterization of the package, with measured quantities shown in Table.~\ref{tab:params_table}. 

We used a $\lambda/4$ cavity made of 4N Aluminum as the storage mode~\cite{reagor_quantum_2016} in which we drive the spin dynamics.  
The cavity was treated with an acid etch to remove surface impurities. 
While this process typically produces cavities with lifetimes approaching the millisecond scale, the surface appears to have degraded due to shipment, resulting in the measured cavity lifetime of \SI{132}{\micro\second}. 
However, after another round of etching the cavity for 2 hours with transene Al etchant at 50C the lifetime has increased to \SI{396}{\micro\second} and we have used this cavity for taking the data in Figures 3, 4, 5 and S13.  

The superconducting qubit used is a transmon with Nb capacitor pads, and an Al/AlO\textsubscript{x}/Al Josephson junction fabricated on a high-resistivity ($>\SI{10}{\kilo\ohm\centi\meter}$) silicon chip. 
Following a strip of the native silicon oxide in 2\% hydrofluoric acid, a \SI{75}{\nano\meter} thick Nb film was sputtered at the rate of \SI{50}{\nano\meter\per\minute}.
This Nb was patterned using photo-lithography to make the capacitor pads of transmon and the readout resonator.    
The Josephson junction for the qubit was fabricated in a Dolan bridge process~\cite{dolan_offset_1977} with bilayer MMA/PMMA resist. 
Double-angle evaporation for the Al-AlO\textsubscript{x}-Al was accomplished to form the junction. Before the first Al deposition an in-situ ion mill was performed to clean the top surface of Nb. After the first Al deposition, AlO\textsubscript{x} layer was formed by oxidizing the Al surface for 30 minutes at 7.25 Torr oxygen pressure. After pumping out the oxygen from the chamber, the second layer of Al was deposited . 
The area of the deposited Josephson junction is \SI{0.047}{\micro\meter^2}.
The chip  with the transmon qubit and readout resonator is mounted in the 3D cavity with a copper clamp .

\begin{figure*}
\centering
\includegraphics[width = 0.85\textwidth]{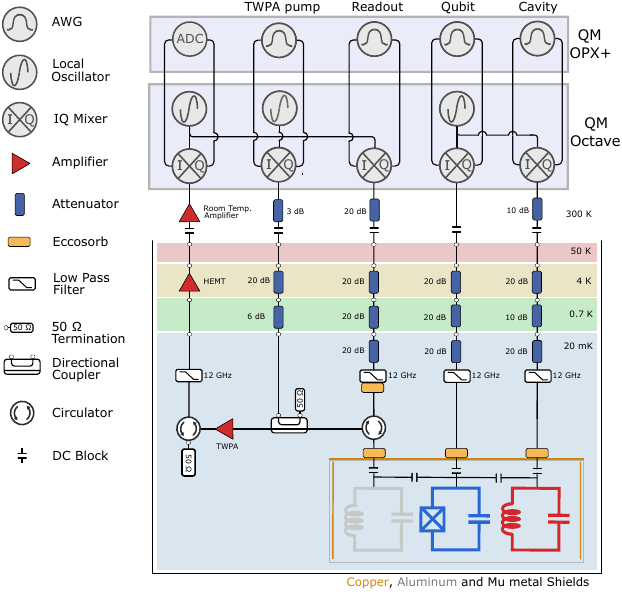}
\caption{\textbf{Experimental wiring diagram}: 
Schematic of room temperature and cryogenic microwave circuit. 
}
\label{fig:sf_wiring}
\end{figure*}

\begin{figure}
\centering
\includegraphics{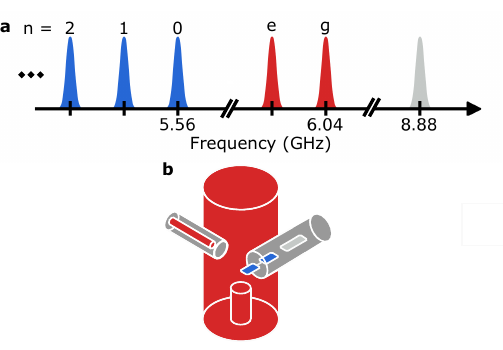}
\caption{
\textbf{Pulses in Frequency domain and device schematic:} 
(a) The MEM comb (blue) on the qubit has 2J+1 frequencies at $\omega_q + n \chi$, $n$ = 0, 1,..,2J . The cavity pulse (red) for driving the spin system has two frequencies at $\omega_c$ and $\omega_c + \chi$ for all spins. 
Readout (gray) is always performed at one frequency.
The diagram indicates the frequencies and the approximate pulse bandwidths, but the shapes themselves should not be taken literally (In time domain, the pulses are flat-top with Gaussian rise and fall of about \SI{150}{\nano\second} for qubit pulses and \SI{400}{\nano\second} for cavity pulses and $\sigma$ = \SI{38}{\nano\second} for qubit and \SI{80}{\nano\second} for cavity pulses).
(b) Schematic of the cavity-qubit experimental device. A coaxial $\lambda/4$ microwave cavity (red) is used as the harmonic oscillator. The qubit (blue) and readout resonator (gray) are on-chip. The chip goes partially into the cavity via a tunnel. A cylindrical coupling pin made of gold plated beryllium copper (shown in red on the side) is used to drive the cavity. Separate pins are used to drive the qubit and the readout (not shown in picture)}

\label{fig:sf_freq}
\end{figure}

The whole package is mounted on an OFHC copper bracket and has multiple layers of shielding, in order from inner to outer: (1) a Berkeley Black \cite{10.1063/1.1149739} coated copper shim to absorb any stray mm-wave radiation, (2) an aluminum can with an indium seal to an upper flange, which itself has indium-sealed SMA feedthroughs, (3) a mixing chamber can. 
This is depicted schematically in Fig.~\ref{fig:sf_wiring}.
The dilution fridge also hosts a room-temperature magnetic shield that lines the vacuum can. 

The control pulses are generated using Quantum Machines OPX plus instrument and up-converted using local oscillator (LO) and IQ mixers in the Quantum Machines Octave. 
All pulses are digitally triggered with \SI{200}{\nano\second} buffer on the trigger on either side of the analog pulse.
The qubit and storage drives share the same LO whereas the readout up and down conversion share the same LO.  
Output of the readout signal is first amplified with a Travelling wave Parametric Amplifier (TWPA) at base temperature, followed by a High electron mobility transistor (HEMT) at \SI{4}{\kelvin} and a room temperature amplifier (ZVA-1W-103+ Mini-Circuits). 
The TWPA is also pulsed with \SI{100}{\nano\second} buffer on both ends of the readout pulse.
Each of the input lines have K \& L low pass filter and Eccosorb filters. 
All the last attenuators, microwave filters, and Eccosorb filters are connected to the 
copper bracket using a copper braid to improve thermalization.

\begin{table*}
    \centering
    \begin{tabular}{ | c | c | c | c | c |}

         \hline
         \textbf{Quantity}  & \textbf{Symbol} & \textbf{Measured value (1)} & \textbf{Measured value (2)} & \textbf{Hamiltonian term}\\
         \hline \hline
         Cavity frequency  & $\omega_c / 2\pi$ & \SI{6.04}{\giga\hertz} & \SI{6.043}{\giga\hertz} & $\omega_c c^ \dagger c$\\
         \hline
         Readout frequency  & $\omega_r / 2\pi$ & \SI{8.88}{\giga\hertz} & \SI{8.92}{\giga\hertz} & $\omega_r r^ \dagger r$\\
         \hline
         Qubit frequency & $\omega_q / 2\pi$ & \SI{5.56}{\giga\hertz} & \SI{5.57}{\giga\hertz} & $\omega_q q^ \dagger q$\\
         \hline
         Qubit-readout dispersive coupling  & $\chi_r / 2\pi$ & \SI{-700}{\kilo\hertz} & \SI{-700}{\kilo\hertz} & $\chi_r r^\dagger r |e\rangle \langle e|$ \\
         \hline
         Qubit-Cavity dispersive coupling  & $\chi/2\pi$ & \SI{-2.54}{\mega\hertz} & \SI{-3.56}{\mega\hertz} & $\chi c^\dagger c |e\rangle \langle e|$\\
         \hline
         Qubit anharmonicity  & $\alpha/2\pi$ & \SI{-180}{\mega\hertz} & \SI{-180}{\mega\hertz} & $\alpha q^\dagger q^\dagger q q $/2\\
         \hline
         Second order cavity-qubit dispersive coupling  & $\chi^\prime/2\pi$ & \SI{6.5}{\kilo\hertz} & \SI{7}{\kilo\hertz} & $\chi^\prime c^\dagger c^\dagger c c |e \rangle \langle e|$/2\\
         \hline
         Cavity Self-Kerr & $K/2\pi$ & \SI{-9}{\kilo\hertz} & \SI{-11}{\kilo\hertz} & $K c^\dagger c^\dagger c c $/2\\
         \hline
         Qubit lifetime  & $T_{1q}$ & \SI{90}{\micro\second} & \SI{90}{\micro\second} & \\
         \hline
         Qubit dephasing time, Ramsey  &  $T_{2q,R}$ & \SI{40}{\micro\second} & \SI{40}{\micro\second} & \\
         \hline
         Qubit dephasing time, Hahn-echo &  $T_{2q,E}$ & \SI{40}{\micro\second} & \SI{38}{\micro\second} & \\
         \hline
         Cavity lifetime   & $T_{1c}$ & \SI{132}{\micro\second} & \SI{396}{\micro\second} & \\
         \hline
         Cavity dephasing time   & $T_{2c,R}$ & \SI{150}{\micro\second} &  \SI{160}{\micro\second} & \\
         \hline         
         Qubit thermal population  &    &  3\% & 2.5\%  & \\
         \hline
         Cavity thermal population   &  & 0.7\% & 0.7\% & \\
         \hline
         
    \end{tabular}
    \caption{System parameters measured using standard time-domain and spectroscopy techniques. 
    The second column of values were measured in a different cool down with a newly etched cavity of the same design. 
    This package was used to obtain the data shown in Fig. \ref{fig:fig3}, \ref{fig:fig_gen_rots}, \ref{fig:fig4}, \ref{fig:sf_spincat_gates_Y}.
    Rest of the article uses data taken with the device with parameters given in the first column of measured values.
    The qubit has some TLS associated with it which makes it responds at two different frequencies about 50-\SI{60}{\kilo\hertz} apart as seen in qubit spectroscopy.
    For determining cavity dephasing time, a Ramsey experiment is performed. 
    The $\pi/2$ pulse used for Ramsey experiment puts the cavity in superposition $(|0\rangle + |1\rangle)/\sqrt{2}$. 
    This $\pi/2$ pulse is achieved by doing a photon Blockade at $n=2$ cavity state and driving the cavity for an appropriate duration.
    }
    \label{tab:params_table}
\end{table*}

\subsection{Calibrating returning the qubit to $| g \rangle$ at end of protocol}

\label{app:technical_calibration}

\begin{figure}
\centering
\includegraphics{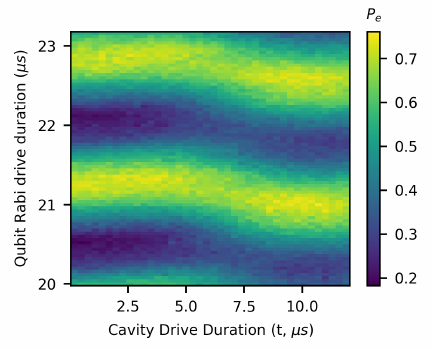}
\caption{\textbf{Checking the qubit state at the end of the protocol}: 
The qubit drive is slightly varied at larger than cavity drive duration to check when the qubit returns to ground state for different duration of cavity drive.
For a given protocol, the cavity is driven for particular duration and a corresponding qubit drive duration is chosen from the map such that qubit returns to the ground state as much as possible.}
\label{fig:sf_background}
\end{figure}

Since we are using the same qubit for modifying the harmonic oscillator to a spin system and using it to check the population of the cavity, it is important to make sure that at the end of our protocol the qubit ends in ground state, so that we can probe the cavity population correctly. 
The calibration to bring the qubit back to ground state at the end is also important so that we can say an error (when the qubit has suffered a $T_1$ or $T_2$ event) has happened when the qubit is found in the excited state at the end of the protocol. 
Then we can post-select on cases where the qubit has ended in ground state and keep most of the experiments for building the statistics of cavity population.
We perform an experiment similar to what is shown in Fig.~\ref{fig:fig2} where we keep the qubit drive duration fixed and vary the cavity drive duration to see the spin dynamics. 
At the end of the spin protocol, qubit is disentangled from the cavity and qubit state is probed. 
We repeat this experiment with slightly different qubit drive duration. 
This gives a map of qubit state at the end of the protocol for different duration of qubit drives and cavity drives. 
Because the duration for which the qubit experiences Stark shifts (due to the cavity drive) is varying (as cavity drive duration is changed), the time required for the qubit to return to ground state also varies slightly. 
This is shown in Fig.~\ref{fig:sf_background}.
For preparing a state (or performing a gate) the cavity drive needs to be turned on for a particular time. 
The corresponding duration of qubit drive is chosen for this length of cavity drive from the color map. 
The qubit drive duration is chosen to be the least such duration which is larger than the maximum cavity drive duration required for a particular protocol to avoid decoherence related errors.

\balancecolsandclearpage

\section{Additional comparisons between HAMS dynamics and photon-blockaded dynamics} 

\subsection{Non-periodic dynamics under standard photon Blockade}\label{app:aperiodic}
The Hamiltonian of a cavity linearly driven on resonance is given by $\hH \propto \hat x$, where we performed a RWA and chose the drive phase without loss of generality. 
Under standard photon blockade~\cite{bretheau_quantum_2015} which limits the maximum photon number to $N$, the matrix elements of a microwave drive remain unmodified, except for the $N \leftrightarrow N+1$ matrix elements which are set to 0. In that context, the Hamiltonian of the photon-blockade driven cavity is $\hH_{\mathrm{FB}} \propto \hat x_J $, where we have defined
\begin{equation}
    \hat x_J \equiv \hPi_J \hat x \hPi_J.
\end{equation}
where $\hPi_J = \sum_{n=0}^N |n\rangle \langle n|$. 
Evolution under that Hamiltonian is given by the unitary operator $\hU = \exp(-i \theta \hat x_J)$, such that a periodic evolution where $\hU(\theta_*) \propto \mathbb I$ implies that the eigenvalues $\{\lambda_j\}$ of $\hat x_J$ are distributed such that $\theta_* \lambda_j = \phi + 2\pi k_j$, where $\phi$ is a resulting global phase and $k_j \in \mathbb Z$.
The eigenvalues of $\hat x_J$ are obtained by finding the zeros its characteristic polynomial, which in turn is given by the determinant of the tri-diagonal matrix
\begin{equation}
    \det(\hat x_J - \lambda \mathbb I) = 2^{-(2J+1)/2} H_{2J+1}\left(\frac{-\lambda}{\sqrt 2}\right),
\end{equation}
where $H_n(x)$ is the $n$th order Hermite polynomial. The eigenvalues of $\hat x_J$ are therefore given by the zeros of the Hermite polynomials. 
Zeros of Hermite polynomials come in pairs $\pm \lambda$ (except for odd-order polymials with an additional $\lambda = 0$ eigenvalue).
Accordingly, we have $\det(\exp(-i \theta \hat x_J)) = 1$ for all $\theta$, which means that the condition on the eigenvalues to obtain periodic evolution is given by $\theta_* \lambda_j = 2\pi k_j$ for some $\theta_*$. 
This means that the ratio between eigenvalues must be a rational number, $\lambda_j/ \lambda_l = k_j/k_l \in \mathbb Q$ for all $\lambda_j,\lambda_l \neq 0$.

While we were not able to formally prove that such ratios were irrational (implying an aperiodic evolution), we numerically checked that these ratios needed at least 18 digits to rationalize for $1 < J \leq 25$, making the evolution generated by $\hat x_S$ aperiodic for all practical purposes. 
For $J=1/2$, the system reduces to an effective qubit and $\hat x_J \sim \hat \sigma_x$, with a periodic evolution. 
For $J=1$, the eigenvalues of $\hat x_J$ are of the form $\{0,\pm \lambda_1\}$, such that the evolution is periodic with a period $T = 2\pi /\lambda_1$.

\begin{figure}
\centering
\includegraphics[width = \columnwidth]{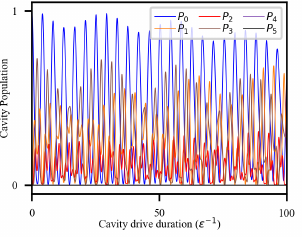}
\caption{\textbf{Standard Fock Blockade dynamics:} 
Simulation of dynamics of the cavity population starting from Fock $|0\rangle$ under the cavity drive while qubit Rabi drive is kept on at the frequency $ \omega_q + 4\chi$. The qubit drive truncates the Hilbert space to the first four levels. The cavity population dynamics is non-periodic for such standard photon blockade. 
The possibility of a long-time periodicity is precluded by the irrational ratios of the rotation generator eigenvalues.
}
\label{fig:sf_blockade}
\end{figure}

As a result, any state with initial support on all eigenstates of $\hat x_J$ (for example, the vacuum state $|0\rangle$) will exhibit non-periodic dynamics under drive. 
We show a simulation of photon blockade at cavity Fock state $|5\rangle$ in Fig.~\ref{fig:sf_blockade} as an example of such non-periodic dynamics. 
The simulation is done for a lossless system with parameters $\epsilon  = 0.01\Omega = 10^{-4}|\chi|$, which satisfies $|\chi|\gg\Omega\gg\epsilon$. 

\subsection{Theoretical and Experimental Comparisons of HAMS and photon blockade dynamics}\label{app:expt_comparisons}

\begin{figure*}
\centering
\includegraphics[width = 0.84\textwidth]{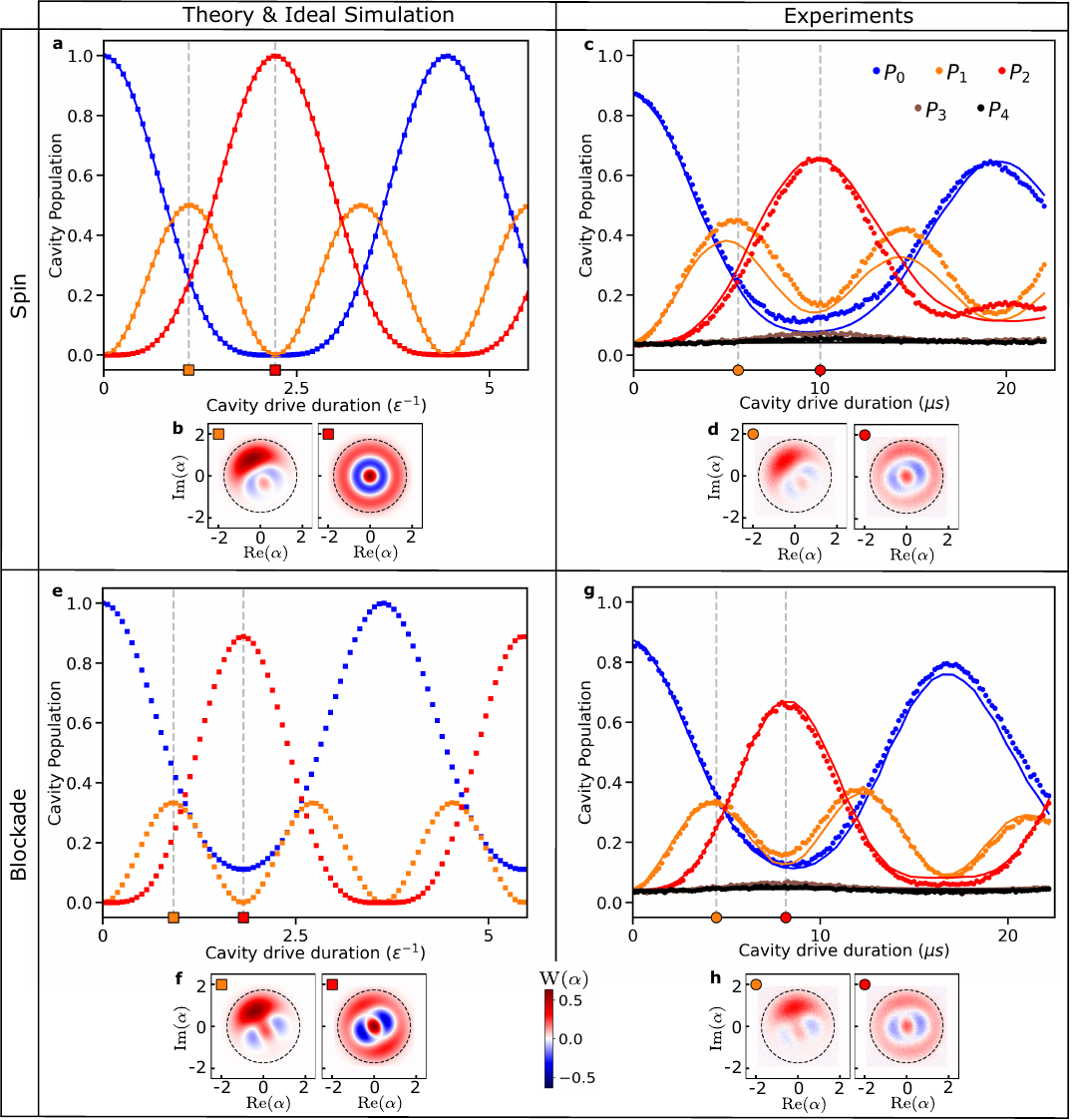}
\caption{\textbf{Spin 1 and 3-level blockade dynamics:} (a) Dynamics of spin states of a driven spin 1 system. 
The points are numerical simulations of our protocol in the ideal limit $|\chi| \gg \Omega \gg \epsilon$.
The solid line is for an ideal driven spin 1, which matches our simulated spin.
(b) Wigner functions of the cavity state for a spin at the marked points are shown. 
(c) Experimentally measured probabilities of different spin states of a driven synthetic spin 1 in our microwave cavity with $\epsilon/2\pi =  \SI{72}{\kilo\hertz}$, $\Omega/2\pi =  \SI{690}{\kilo\hertz}$, $\chi / 2\pi =  \SI{-2.54}{\mega\hertz}$. 
The experimental data is shown in dots and the solid lines are numerical simulation of the spin protocol that consider experimental non-idealities and decoherences. 
(d) Experimentally measured Wigner functions at the marked critical points. 
(e) Dynamics of cavity states when the first three levels of a harmonic oscillator is blockaded without any phase tuning on qubit drive only to limit the Hilbert space dimension to 3. 
(f) Wigner functions of the cavity state at the marked points for a three level blockade are shown.
(g) Experimentally measured probabilities of cavity levels for blockaded dynamics (dots) and simulated dynamics (solid lines) considering experimental non-idealities and decoherences. 
(h) Experimentally measured Wigner functions at the marked points for a three level blockade.
}
\label{fig:sf_spin1}
\end{figure*}

\begin{figure*}
\centering
\includegraphics[width = 0.76\textwidth]{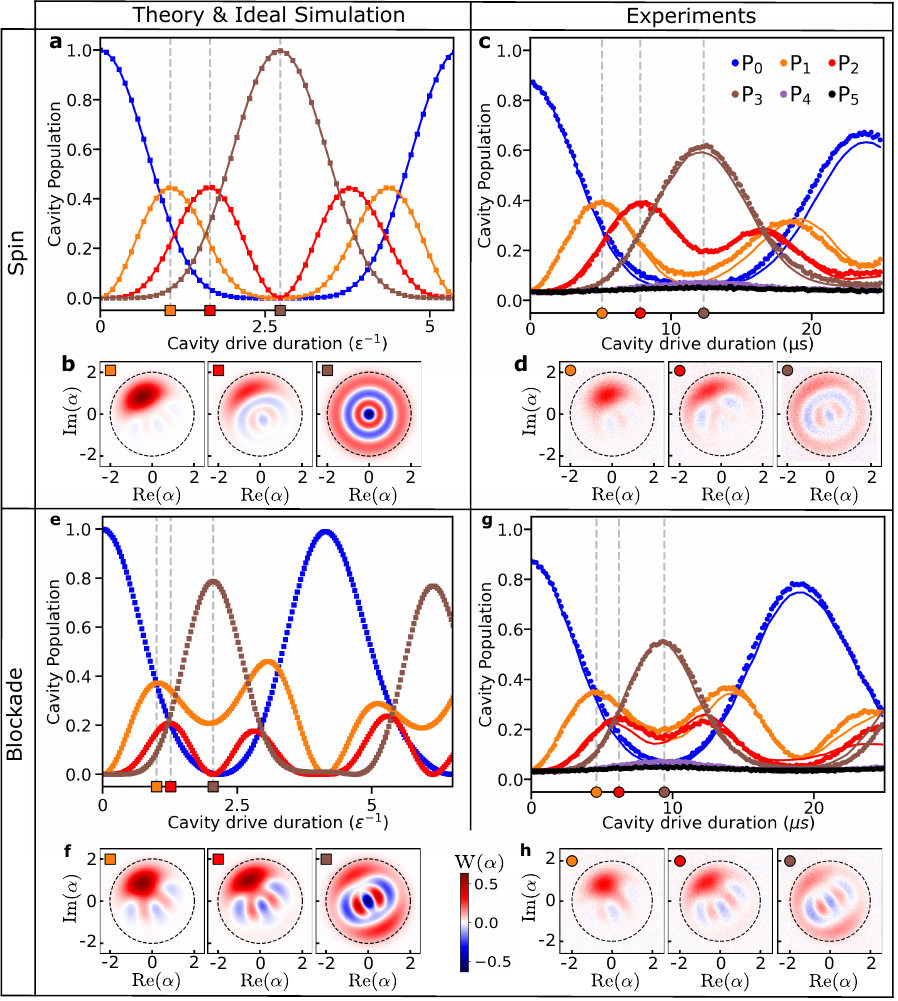}
\caption{\textbf{Spin 3/2 and 4-level blockade dynamics:} (a) Dynamics of spin states of a driven spin 3/2 system. 
The points are numerical simulations of our protocol in the ideal limit $|\chi| \gg \Omega \gg \epsilon$.
The solid line is for a real driven spin 3/2, which matches our simulated spin.
(b) Wigner functions of the cavity state for a spin at the marked points are shown. 
(c) Experimentally measured probabilities of different spin states of a driven synthetic spin 3/2 in our microwave cavity with $\epsilon/2\pi =  \SI{72}{\kilo\hertz}$, $\Omega/2\pi =  \SI{690}{\kilo\hertz}$, $\chi / 2\pi =  \SI{-2.54}{\mega\hertz}$. 
The experimental data is shown in dots and the solid lines are numerical simulation of the spin protocol that considers experimental non-idealities and decoherences. 
(d) Experimentally measured Wigner functions at the marked critical points. 
(e) Dynamics of cavity states when the first four levels of a harmonic oscillator is blockaded without any phase tuning on qubit drive only to limit the Hilbert space dimension to 4. 
(f) Wigner functions of the cavity state at the marked points for a four level blockade are shown.
(g) Experimentally measured probabilities of cavity levels for blockaded dynamics (dots) and simulated dynamics (solid lines) considering experimental non-idealities and decoherences. 
(h) Experimentally measured Wigner functions at the marked points for a four level blockade.}
\label{fig:sf_spin1.5}
\end{figure*}

\begin{figure*}
\centering
\includegraphics[width = 0.85\textwidth]{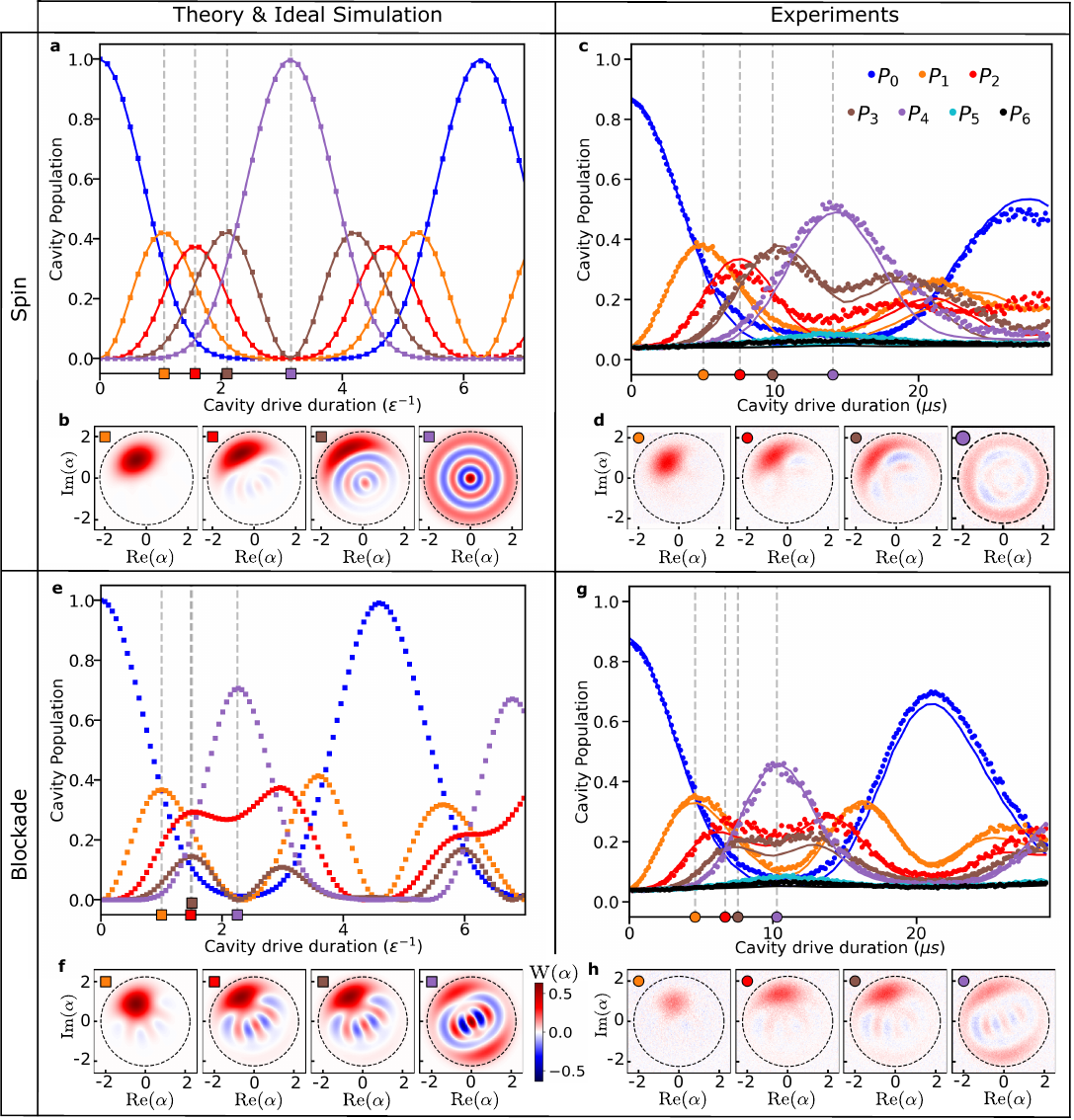}
\caption{\textbf{Spin 2 and 5-level blockade dynamics:} (a) Dynamics of spin states of a driven spin 2 system. 
The points are numerical simulations of our protocol in the ideal limit $|\chi| \gg \Omega \gg \epsilon$.
The solid line is for a real driven spin 2, which matches our simulated spin.
(b) Wigner functions of the cavity state for a spin at the marked points are shown. 
(c) Experimentally measured probabilities of different spin states of a driven synthetic spin 2 in our microwave cavity with $\epsilon/2\pi =  \SI{72}{\kilo\hertz}$, $\Omega/2\pi =  \SI{690}{\kilo\hertz}$, $\chi / 2\pi =  \SI{-2.54}{\mega\hertz}$. 
The experimental data is shown in dots and the solid lines are numerical simulation of the spin protocol that considers experimental non-idealities and decoherences. 
(d) Experimentally measured Wigner functions at the marked critical points. 
(e) Dynamics of cavity states when the first five levels of a harmonic oscillator is blockaded without any phase tuning on qubit drive only to limit the Hilbert space dimension to 5. 
(f) Wigner functions of the cavity state at the marked points for a five level blockade are shown.
(g) Experimentally measured probabilities of cavity levels for blockaded dynamics (dots) and simulated dynamics (solid lines) considering experimental non-idealities and decoherences. 
(h) Experimentally measured Wigner functions at the marked points for a five level blockade.}
\label{fig:sf_spin2}
\end{figure*}

For a spin 1/2 system we do not have to modify the phase difference of the Rabi drives at the frequencies $\omega_q$ and $\omega_q + \chi$. 
For HAMSs, the phase of the Rabi drives at $\omega_q + n\chi$ with $n = 2, 3,..., 2J$, need to be changed to modify the matrix elements of the oscillator, so that the driven system can behave as a HAMS.
Figs. \ref{fig:sf_spin1}, \ref{fig:sf_spin1.5}, and \ref{fig:sf_spin2} compare theory, simulations, and experiments for the case of simple blockade (where the qubit drive phases have no relative difference to each other, equivalent to previous work ~\cite{bretheau_quantum_2015}) and HAMS where the qubit drives have appropriate Fock state dependent phases according to Eq.~\eqref{eqn:phase_modification}. 

Fig.~\ref{fig:sf_spin1} shows the case of spin 1 and blockade of first 3 levels of the harmonic oscillator.  
Eq.~\eqref{eqn:phase_modification} provides the phases $\phi_0 = \phi_1 = 0$, $\phi_2 = \pi/2$. 
For this modification a lossless ideal cavity-qubit system obeying the theoretical limit $|\chi| \gg \Omega \gg \epsilon$ can accurately simulate the dynamics of a HAMS.
Fig.~\ref{fig:sf_spin1}(a) shows the cavity population dynamics with dots for a cavity-qubit setup and real spin dynamics in solid lines. 
Fig.~\ref{fig:sf_spin1}(b) shows the dynamics of cavity-qubit system that is observed in experiment (dots). 
The dynamics is captured well in our simulations (solid lines) that take into account the experimental parameters and decoherence.

Without any drive phase differences, i.e. $\phi_0 = \phi_1 = \phi_2 = 0$, we achieve a usual blockade: the oscillator is truncated to first three levels but it shows a dynamics different than a spin. 
To highlight the distinction, we inspect the cavity state at the quarter period.
We make the same comparisons for the case of spin 3/2 and spin 2 in Figs. \ref{fig:sf_spin1.5} and \ref{fig:sf_spin2}, respectively. 
Note that the measured Wigner functions of the state at ``quarter period" (note that dynamics for simple blockade here are aperiodic) further show the distinction between spin and truncated oscillator dynamics. 
The clarity of the difference between usual blockade and SU(2) dynamics improves as the size of the manifold gets larger, in both the population dynamics and the Wigner functions at the sampled points.
For creating spin 3/2, the phases are $\phi_0 = \phi_1 = 0$, $\phi_2 = 1.231$, $\phi_3 = \pi$. 
For creating spin 2 the phases are $\phi_0 = \phi_1 = 0$, $\phi_2 = 1.047$, $\phi_3 = 2.618$, $\phi_4 = 4.712$. 

\subsection{Effect of non-adiabaticity and decoherence}
\label{app:non-adiabaticity}

The coherent errors in the experiment were dominated by the violation of strict adiabatic limit.  
In Table~\ref{app_tab:error_budget}, we show the contribution of various losses present in our system. 
We show the highest population in Fock states $|3\rangle$ and $|4\rangle$ for a driven spin 3/2 in our cavity, when different losses are considered. 
$P_3$ indicates the fidelity of Fock state $|3\rangle$ preparation using spin rotations and $P_4$ indicates leakage out of the spin manifold. 
Most of the loss can be attributed to qubit $T_1$ and the protocol is presently only weakly affected by qubit $T_2$. 
Cavity self-Kerr and second order dispersive shifts are also non-idealities present in our system which we have not yet separately considered. 
Apart from the error due to loss, we will also have errors because of violation of adiabatic limit, self-Kerr, second order dispersive shift and other higher order terms.

\par

Previously we mentioned in Appendix \ref{app:methods}, that a maximum of 35\% of shots are rejected for spin evolution experiments and about 15\% shots are rejected for spin cat gates experiments because of the qubit not returning to ground state. 
For the specific duration of the drive we expect the qubit will be found in excited state 25\% and 15\% of the time respectively because of $T_1$ errors. 
At this time, we attribute the rest of the probability for qubit not ending in ground state to non-adiabaticity and Stark shift due to presence of multiple drives.

\begin{table*}
    \centering
    \begin{tabular}{ | c | c | c | c | c | c|}

         \hline
         \textbf{Qubit $T_1$}  & \textbf{Qubit $T_2$} & \textbf{Cavity $T_1$} & \textbf{Cavity $T_2$} & \textbf{Max($P_3)$} & \textbf{Max($P_4$)}\\
         \hline \hline
         \tikzcmark  & \tikzcmark & \tikzcmark & \tikzcmark & 0.736 & 0.037 \\
         \hline
         \tikzcmark &  &  &  & 0.79 & 0.04\\
         \hline
          & \tikzcmark &  &  & 0.83 & 0.04\\
         \hline
          &  & \tikzcmark &  & 0.83 & 0.04\\
         \hline
           &  &  & \tikzcmark & 0.85 & 0.04\\
         \hline
          &  &  &  &  0.86 & 0.04 \\
         \hline
         
    \end{tabular}
    \caption{The table lists the highest population in Fock states $|3\rangle$ and $|4\rangle$ for a driven spin 3/2 system hosted in the cavity while different losses of the system are considered (checks indicate which decoherence time was included in the simulation of that row). 
    Maximum of $P_3$ and $P_4$ respectively signify the fidelity of Fock state preparation and leakage out of the spin manifold. 
    Here we observe that the highest error occurs because of the qubit $T_1$, while the effect of qubit and cavity $T_2$ is comparatively small. 
    The simulations use $\chi=$\SI{3.56}{\mega\hertz}, $\Omega=$\SI{0.732}{\mega\hertz}, $\epsilon=$\SI{80}{\kilo\hertz} which are the parameters used for spin 3/2 dynamics experiment in Fig.~\ref{fig:fig3}.
    }
    \label{app_tab:error_budget}
\end{table*}

\balancecolsandclearpage

\section{Spin cats} \label{app:spincat}

\subsection{Experimental creation of the $S=1$ spin kitten}
\label{app:spincat_prep}
\begin{figure}
\centering
\includegraphics{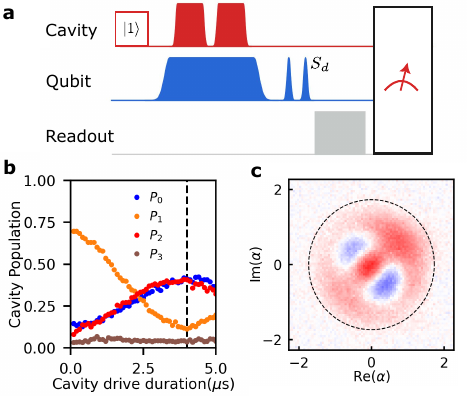}
\caption{\textbf{Spin cat state creation and gates}: (a) Pulse schematic for creation of a spin cat state. Fock state $|1\rangle$ is created in cavity using conventional photon blockade, and then a spin 1 $\hat{J}_y$ drive creates the spin cat state $|0\rangle - |2\rangle$. The next cavity drive is frame rotated ($\phi$) and a particular angle ($\theta$) rotation is done to accomplish the spin cat gate. (b) Cavity population when the $\hat{J}_y$ drive is turned on after creating Fock state $|1\rangle$. (c) Measured Wigner function of the created spin cat state at the \SI{4}{\micro\second} point marked in (b). }
\label{fig:sf_spincat}
\end{figure}

Here we show an interesting way of creating spin kitten state, which was implemented in the experiments. 
First, we use conventional photon blockade \cite{bretheau_quantum_2015} to create nearly pure Fock state $|1\rangle$ in the cavity. 
In this process we limit the Hilbert space of cavity to two energy levels by driving the qubit at frequency $\omega_q + 2\chi$. 
While the Hilbert space is truncated, an appropriate duration of cavity drive creates Fock state $|1\rangle$ in the cavity. 
Then we use our spin displacement protocol to accomplish a spin 1 $\hat{J}_y$ rotation of angle $\pi/2$ on the cavity.
This operation creates the spin kitten state $|0\rangle - |2\rangle$ as shown in Fig.~\ref{fig:sf_spincat}(b) and (c).
Equivalently a $\hat{J}_x$ rotation on the cavity would also create a spin kitten state which is rotated by 90 degrees in phase space i.e. the state -$i$($|0\rangle + |2\rangle$).

For creating the $|\pm Y\rangle$ states of the spin cat, we use superposition of $\hat{J}_x$ and $\hat{J}_y$ operator.

Once the spin kitten state is created we can perform different gates on it, as described in Fig.~\ref{fig:fig4} and associated text. 

\subsection{Logical operations for two-legged spin cats}
The natural spin cat generalization of our $\ket{0} \pm \ket{2}$ encoding are the polar 2-legged spin cats (i.e. code spaces defined via equal superpositions of 2 spin coherent states located at the poles, which are also the states $\ket{-J}$ and $\ket{J}$) given by
\begin{equation}
    \begin{aligned}
        \ket{0_L} &:= \frac{\ket{0} + \ket{N}}{\sqrt{2}} \leftrightarrow \frac{\ket{-J} + \ket{J}}{\sqrt{2}} \\
        \ket{1_L} &:= \frac{\ket{0} - \ket{N}}{\sqrt{2}} \leftrightarrow \frac{\ket{-J} - \ket{J}}{\sqrt{2}}.
    \end{aligned}
\end{equation}
In these 2-legged spin cats, up to $N - 1$ photon losses, $\{\hat{a}^{n}\hspace{2pt}|\hspace{2pt}0 < n < N\}$, can be detected (but not corrected) since these errors bring both code words outside of the code space: $\hat{a}^{n}|\mu_{L}\rangle \propto |N - n\rangle$ for $\mu \in \{0, 1\}$.
\par
We are particularly interested in the logical operations for these general 2-legged spin cats that can be achieved via SU(2) rotations of the synthetic spin $J = \frac{N}{2}$ system.
By looking at the action of $\hat{R}_{z}^{(J)}(\theta')$ spin rotations on the $\ket{\pm_L} = \ket{\mp J}$ logical states, we can first note that these spin rotations yield logical operations corresponding to any Bloch sphere rotation about the $x$-axis:
\begin{equation}
    \begin{aligned}
        \hat{R}_z^{(J)}(\theta')\ket{\pm_L} &= \expo{-i\theta' \hat{J}_z^{(J)}}\ket{\mp J} \\
        &= \expo{-i\theta' (\mp J)}\ket{\mp J} \\
        &= \expo{-i\frac{-2\theta' J}{2}\hat{X}}\ket{\pm_L} \\
        &= \hat{R}_x^L(-2\theta' J)\ket{\pm_L}.
    \end{aligned}
\end{equation}
Since any logical unitary is completely specified by its action on two distinct logical states, we thus indeed see that any given Bloch sphere rotation $\hat{R}_x^L(\theta)$ is achieved by any of the $2J$ distinct SU(2) rotations $\hat{R}_z^{(J)}\left(-\frac{\theta + 2\pi k}{2J}\right)$ (where $k$ is an integer such that $0 \leq k< 2J$).
Setting $\theta = 0$, this also implies that there are always $2J - 1$ nontrivial SU(2) rotations that act as stabilizers on the code space.

We remark that all of these $z$-axis SU(2) rotations can be realized with simple phase space rotations $\mathrm{exp}\left(\frac{i(\theta +2\pi k)\hat{n}}{2J}\right)$ (i.e. updates of the local oscillator phase), so these specific logical unitaries can be implemented without the use of SU(2) rotations.
However, using the Wigner D-matrix identity~\cite{wigner_gruppentheorie_1931}
\begin{equation}
    \bra{J, m'}\expo{-i\pi \hat{J}_x^{(J)}}\ket{J, m} = (-1)^{3J}\delta_{m', -m}
\end{equation}
to understand the action of the $\hat{R}_x^{(J)}(\pi)$ on the $\ket{\pm_L}$ logical states, we see that (up to a global phase) this $\hat{R}_x^{(J)}(\pi)$ rotation always yields the logical operation $\hat{Z} = i\hat{R}_z^L(\pi)$, 
\begin{equation}
    \begin{aligned}
        \hat{R}_x^{(J)}(\pi)\ket{\pm_L} &= \expo{-i\pi \hat{J}_x^{(J)}}\ket{\mp J} \\
        &= (-1)^{3J}\ket{\pm J} \\
        &= (-1)^{3J}\ket{\mp_L} \\
        &= (-1)^{3J}\hat{Z}\ket{\pm_L},
    \end{aligned}
\end{equation}
which notably cannot be realized using just phase space rotations.

Since, like logical operations, SU(2) rotations form a group under operation composition, any logical unitary that is a combination of $\hat{Z}$ and $\hat{R}_x^L(\theta)$ -- namely any Bloch sphere rotation by $\pi$ about an axis on the $yz$ great circle -- can also be achieved by an SU(2) rotation. Specifically, the logical operation $\hat{R}_{\sin(\phi)y + \cos(\phi)z}^L(\pi)$ is achieved by any of the 2J distinct SU(2) rotations $\hat{R}_{\cos(\phi')x + \sin(\phi')y}^{(J)}(\pi)$, where $\phi' = \frac{\phi + \pi k}{2J}$ and $k$ is an integer such that $0 \leq k < 2J$.

Based on further investigations we have done (which we will discuss in a later work) regarding the rotation gates of the binomial codes -- which are rotated versions of the polar 2-legged spin cat codespaces when the binomial code has distance 0 to photon loss  -- we believe that these logical unitaries generated by $\hat{Z}$ and $\hat{R}_x^L(\theta)$ constitute an exhaustive list of the logical unitaries achievable via SU(2) rotations. 
In Table.~\ref{tab:spinCatLogicalOps}, we summarize these achievable unitaries for the polar cats discussed here, along with the specific instances of these logical unitaries that generalize (to all spins) the operations experimentally demonstrated in Fig.~\ref{fig:fig4}.

\begin{table*}
    \centering
    \begin{tabular}{| c | c | c | c |}

         \hline
         \textbf{Logical Gate}  & \textbf{Bloch Sphere Rotation}  & \textbf{Spin Rotation} & \textbf{Type} \\
         \hline \hline
         Type I & $\hat{R}_x^L(\theta) = \expo{-i\frac{\theta}{2}\hat{X}}$ & $\hat{R}_z^{(J)}\left(-\frac{\theta + 2\pi k}{2J}\right) = \expo{i\frac{\theta + 2\pi k}{2J}\hat{J}_z^{(J)}}$ & I \\
         \hline
         Type II & $\hat{R}_{\sin(\phi)y + \cos(\phi)z}^L(\pi) = \expo{-i\frac{\pi}{2}\left(\sin(\phi)\hat{Y} + \cos(\phi)\hat{Z}\right)}$ & $\hat{R}_{\cos(\phi')x + \sin(\phi')y}^{(J)}(\pi) = \expo{-i\pi\left(\cos(\phi')\hat{J}_x^{(J)} + \sin(\phi')\hat{J}_y^{(J)}\right)}$ & II \\
         \hline \hline
         
         X & $\hat R_x^L(\pi) = \expo{-i\frac{\pi}{2} \hat{X}}$ & $\hat{R}_z^{(J)}\left(-\frac{\pi}{2J}\right) = \expo{i\frac{\pi}{2J} \hat J_z^{(J)}}$ & I ($\theta = \pi$) \\
         \hline
         Y & $\hat{R}_y^L(\pi) = \expo{-i\frac{\pi}{2} \hat{Y}}$ & $\hat{R}_{\cos\left(\frac{\pi}{4J}\right)x + \sin\left(\frac{\pi}{4J}\right)y}^{(J)}(\pi) = \expo{-i\pi\left(\cos\left(\frac{\pi}{4J}\right)\hat{J}_x^{(J)} + \sin\left(\frac{\pi}{4J}\right)\hat{J}_y^{(J)}\right)}$ & II ($\phi = \frac{\pi}{2}$) \\
         \hline
         Z & $\hat{R}_z^L(\pi) = \expo{-i\frac{\pi}{2} \hat{Z}}$ & $\hat{R}_x^{(J)}(\pi) = \expo{-i\pi\hat{J}_x^{(J)}}$ & II ($\phi = 0$) \\
         \hline
         S$^{\dag}$HS & $\hat{R}_{y - z}^L(\pi) = \expo{-i\frac{\pi}{2\sqrt{2}}\left(\hat{Y} - \hat{Z}\right)}$ & $\hat{R}_{\cos\left(\frac{3\pi}{8J}\right)x + \sin\left(\frac{3\pi}{8J}\right)y}^{(J)}(\pi) = \expo{-i\pi\left(\cos\left(\frac{3\pi}{8J}\right)\hat{J}_x^{(J)} + \sin\left(\frac{3\pi}{8J}\right)\hat{J}_y^{(J)}\right)}$ & II ($\phi = \frac{3\pi}{4}$) \\
         \hline
    \end{tabular}
    \caption{Summary of the polar 2-legged spin cat logical unitaries that can be implemented as spin $J$ SU(2) rotations. The first section lists the two general types of SU(2)-implementable Bloch sphere rotations, with the Type I rotations being those achievable via some phase space rotation and the Type II rotations being those achievable via a $\hat{R}_x^{(J)}(\pi)$ rotation composed with some phase space rotation. Note that in the Type II rotation row, $\phi' = \frac{\phi + \pi k}{2J}$. The second section lists particular ($k = 0$) instances of these types of logical unitaries that correspond to the Pauli gates and the rotated Hadamard gate for general spin. Note that the spin rotations in this second section are not immediate generalizations of the rotations from Fig.~\ref{fig:fig4} since for those experimental demonstrations of the spin 1 case we happened to choose $k = 1$ for the X, Z, and S$^{\dag}$HS gates instead of $k = 0$. Also note that in the gate and rotation columns we have removed the global phase factors.
    }
    \label{tab:spinCatLogicalOps}
\end{table*}

\balancecolsandclearpage

\subsection{Gates on spin cat states:}
\label{app:spin_cat_gate_all}

\begin{figure*}[p]
\centering
\includegraphics[width = 0.9\textwidth]{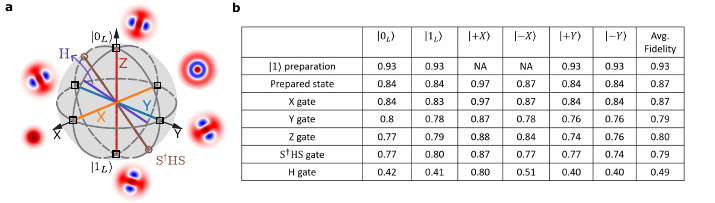}
\caption{\textbf{Gates on spin 1 spin cat:}
(a) The spin 1 cat logical Bloch sphere, with the cardinal points marked with black square boxes and the corresponding Wigner function of the state shown by the side. 
The colored lines show the axis of rotation for the different gate operations as marked.
The Table in (b) reports the fidelity of prepared states before and after different gate operations with ideally expected states. 
We have reconstructed the cavity state from experimentally measured Wigner functions shown in Fig. \ref{fig:sf_spincat_gates_wigners} and calculated the fidelity with the ideally expected state. 
Note that the Wigner function measurement is post-selected on cases where the qubit ends in the ground state at the end of the protocol.}

\label{fig:sf_spincat_gates_Y}
\end{figure*}

\begin{figure*}[p]
\centering
\includegraphics[width = 0.9\textwidth]{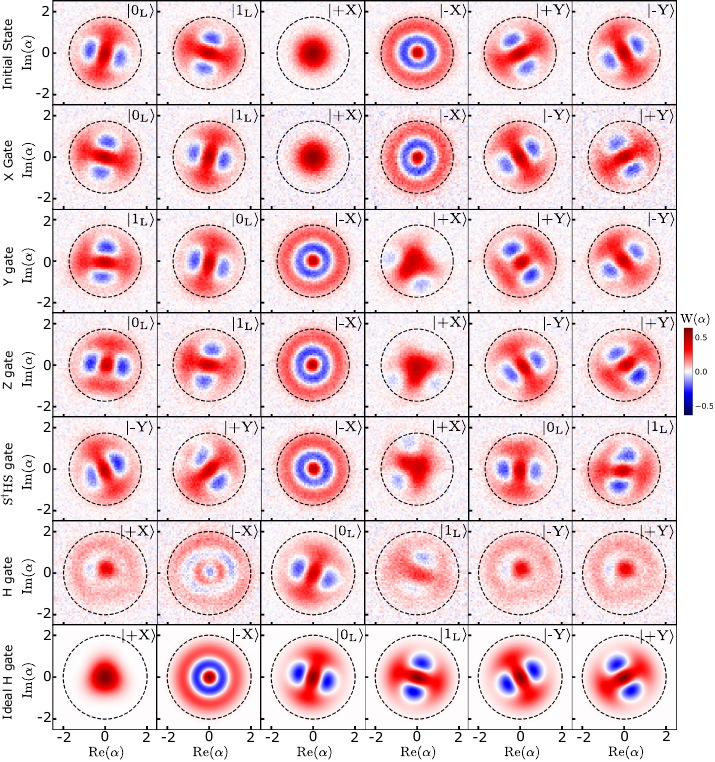}
\caption{\textbf{Measured Wigner function for Gates on spin 1 spin cat:}
Experimentally measured Wigner functions after the state preparation and the application of different gate operations. These Wigner functions have been used to calculate the fidelities shown in the table in b. This shows that the SU(2) rotations generated in our system can accomplish $X$, $Y$, $Z$ and $S^\dagger H S$ gate on all cardinal points of the Bloch sphere. The Wigner functions show that using the nonlinear rotation it is possible to achieve $H$ gate. Since the experimental Wigners are not great due to decoherence issues, we have added an additional row with theoretical Wigners in a lossless system in adiabatic limit ($\epsilon : \Omega : |\chi| = 0.01:1:100$) that shows the accomplishment of H gate. The black circles representing the photon number allowed by the protocol is bigger in radius for the nonlinear rotations as it allows one more energy level to be occupied. 
}

\label{fig:sf_spincat_gates_wigners}
\end{figure*}

In the main text, we discussed how different spin rotations can be used to perform gates on spin 1 cat states.
To provide a more complete demonstration of the gates, here we show that the spin rotations indeed perform the expected gate on all 6 cardinal points of the spin cat logical Bloch sphere as shown in Fig.~\ref{fig:sf_spincat_gates_Y}(a). 
In the table in Fig.~\ref{fig:sf_spincat_gates_Y}(b) we report the fidelities of states in cavity with the theoretically expected states before and after the gates. 
Note that the fidelity is for a state after post-selection of the experiments where qubit ends in ground state at the end of the protocol.  
In experiments, we measure the Wigner function of the cavity state, and reconstruct the state using methods described in \cite{chakram_multimode_2022, PhysRevLett.108.070502} to calculate the fidelity.  
In Fig.~\ref{fig:sf_spincat_gates_wigners}, we show the experimentally measured Wigner functions before and after applying the spin rotation corresponding to all the gate operations. It shows that the four gate operations using linear rotations are accomplished on all six cardinal points on the Bloch sphere. We also show the effect of nonlinear spin rotation on the six cardinal points of the logical Bloch sphere. We have also included a row of simulated Wigner function for this nonlinear rotation. This shows that this nonlinear rotation can indeed accomplish the H gate.

\par
For state preparation we use a combination of conventional photon blockade \cite{bretheau_quantum_2015} and spin rotations. 
For preparing $|0_L\rangle$, $|1_L\rangle$, and $|\pm Y_L\rangle$, we use conventional photon blockade to create $|1\rangle$ followed by a $\pi/2$ spin rotation (see Appendix \ref{app:spincat_prep} for details).
Preparation of $|1\rangle$ using conventional blockade is not perfect and reaches fidelity of 0.93.
After that the spin rotations suffer non-idealities because of not being in strictly adiabatic limit and decoherence also affects the state preparation.
Infidelities from these two steps account for state preparation infidelity.
After the gate operation, the fidelities on average reduce by 0.07, indicating that we can expect gate fidelities of 0.93 or higher for the gates based on SU(2) rotations.

\par
For the nonlinear rotation that accomplishes H gate, fidelities are worse, largely because of the longer duration of the gate which causes more decoherence in the system. 
In future efforts, we will look into using numerical optimization techniques to make higher fidelity and faster gates.  
We note that the gates accomplished by linear rotation have a gate duration of \SI{6.1}{\micro\second} ($2.2 \epsilon^{-1}$) and the gate accomplished by nonlinear rotation has gate duration \SI{19.8}{\micro\second} ($3.8 \epsilon^{-1}$). 

\section{Additional experiments with SU(2) drives}
\label{app:additional}

\subsection{Spin locking}

We can choose the (global) cavity drive phase $\varphi$ to interpolate between $\hat{J}_x^{(J)}$ and $\hat{J}_y^{(J)}$ which means we can change the rotation axis on a high angular momentum spin Bloch sphere.
If the spin is driven for quarter of a period of its dynamics by the $\hat{J}_x$ drive we will create an eigenstate of $\hat{J}_y$. 
If we turn on a $\hat{J}_y$ drive, such a spin state ideally should not change. 

\begin{figure*}
\centering
\includegraphics[width = 0.9\textwidth]{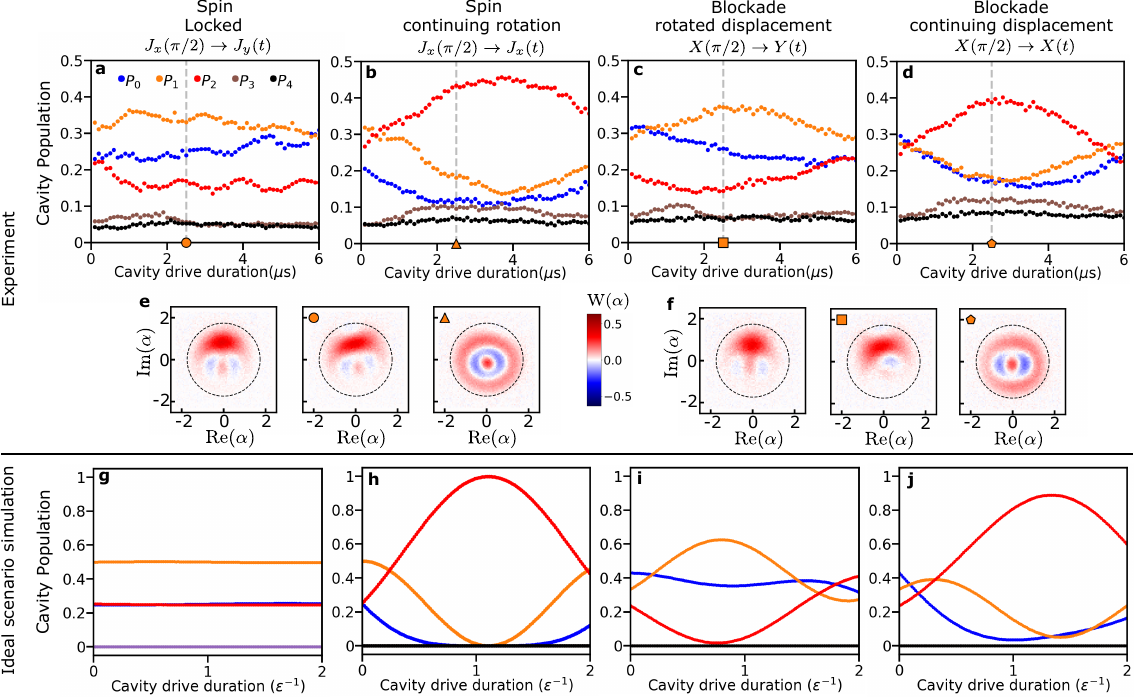}
\caption{\textbf{Spin Locking, spin 1:} We initialize a state on the equator of a spin 1 Bloch sphere by applying a $\hat{J}_x$ rotation and then (a) apply a $\hat{J}_y$ rotation and the cavity population is probed at different duration of the $\hat{J}_y$ drive. Since the created state is on the $\hat{J}_y$ axis, ideally the population should not change. (b) apply a $\hat{J}_x$ rotation and probe the population at different duration of the $\hat{J}_x$ drive. For a simple 3-level blockade, we can also prepare the state at one quarter of the time when population in Fock state 0 returns to maximum, and then (c) drive it along an axis that is rotated by 90 degrees or, (d) drive it along the same axis to probe the population at different times. (e) Experimentally measured Wigner functions for a spin state created on the equator and the states created after \SI{2.4}{\micro \second} of $\hat{J}_y$ and $\hat{J}_x$ drive respectively. (f) Experimentally measured Wigner functions for a created on the equator using 3-level blockade and the states created after \SI{2.4}{\micro \second} of drives on a 90 degree rotated axis and the original axis. (g) - (j) show the simulated probabilities of different cavity states for the cases of spin locking, continued rotation, rotated displacement, and continued displacement (corresponding to experiments in (a)- (d)) achieved by our protocol in an ideal scenario. 
}
\label{fig:sf_spinlock_spin1}
\end{figure*}

We show such a spin locking phenomena in Fig.~\ref{fig:sf_spinlock_spin1}. Since the experiment suffers from non-idealities and decoherence, we compare our spin locking with the case where the $\hat{J}_x$ drive is on for same duration. 
In Fig.~\ref{fig:sf_spinlock_spin1}(e), the measured Wigner function for the created state after applying the $\hat{J}_x$ drive for quarter period and the spin locked state look similar whereas the state created by $\hat{J}_x$ driving is different.

\begin{figure*}
\centering
\includegraphics[width = 1\textwidth]{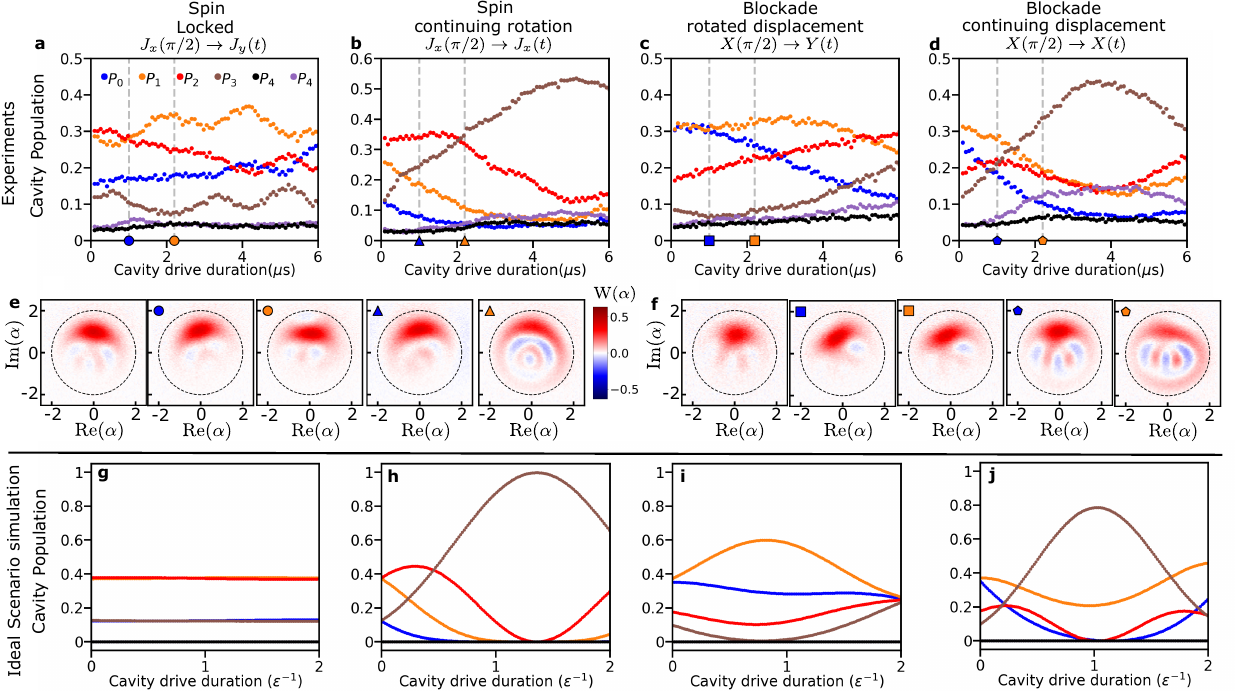}
\caption{\textbf{Spin Locking spin 3/2:} Create a state on the equator of a spin 3/2 Bloch sphere by applying a $\hat{J}_x$ rotation and then (a) apply a $\hat{J}_y$ rotation and the cavity population is probed at different duration of the $\hat{J}_y$ drive. Since the created state is on the $\hat{J}_y$ axis, ideally the population should not change. (b) apply a $\hat{J}_x$ rotation and probe the population at different duration of the $\hat{J}_x$ drive. For a simple 3-level blockade, we can also prepare the state at one quarter of the time when population in Fock state 0 returns to maximum, and then (c) drive it along an axis that is rotated by 90 degrees or, (d) drive it along the same axis to probe the population at different times. (e) Experimentally measured Wigner functions for a spin state created on the equator and the states created after \SI{1}{\micro \second}  and \SI{2.2}{\micro \second} of $\hat{J}_y$ and $\hat{J}_x$ drive respectively. (f) Experimentally measured Wigner functions for a created on the equator using 3-level blockade and 
the states created after \SI{1}{\micro \second}  and \SI{2.2}{\micro \second} of drives on a 90 degree rotated axis and the original axis. (g) - (j) show the simulated probabilities of different cavity states for the cases of spin locking, continued rotation, rotated displacement, and continued displacement (corresponding to experiments in (a)- (d)) achieved by our protocol in an ideal scenario..
}
\label{fig:sf_spin1.5_spinlock}
\end{figure*}

This spin locking phenomenon is also special to spin which is enabled by our phase tuning. 
Simply truncating the Hilbert space by using the frequency comb on the qubit and then changing the axis of rotation does not create such a locking phenomena.
This can be seen in the cavity population and measured Wigner functions shown in Fig.~\ref{fig:sf_spinlock_spin1}(c, d, f).
Since, our protocol suffers from the experimental non-idealities we also show the simulated population dynamics in an ideal system where the difference between spin locking and rotated displacement for blockade is evident. 
For spin 3/2, difference of photon blockade and spin lock becomes clearer. We show experimentally measured Wigner functions at two different times (Fig.~\ref{fig:sf_spin1.5_spinlock}(e)) to show the difference between spin locking and continued rotation.
This is accompanied by an ideal simulation of the spin and photon blockade dynamics. 

\subsection{Detuned Spin Rotations (Spin Chevrons)}

\begin{figure*}
\centering
\includegraphics[width = \textwidth]{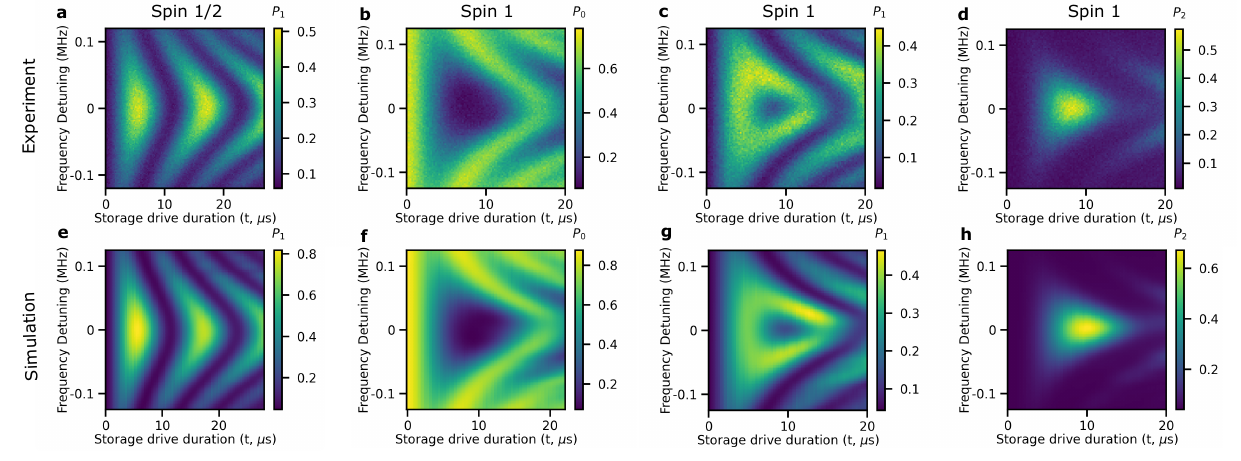}
\caption{\textbf{Spin Chevron:} Both cavity drives for spin coherent rotation are detuned to create a chevron-like plot for spins. Spin Chevron plots for (a) spin 1/2, probing the population of Fock state $|1\rangle$, (b) spin 1, probing the population of Fock state $|0\rangle$. (c)  spin 1, probing the population of Fock state $|1\rangle$, (d) spin 1, probing the population of Fock state $|2\rangle$. e-h are corresponding simulations using experimental parameters. 
}
\label{fig:sf_chevron}
\end{figure*}

For two-level qubit system, the qubit drive frequency can be slightly detuned in a time Rabi experiment, to create a qubit chevron experiment where we can see the Rabi rate changing with detuning. 
For spins created in our cavity we can also detune the cavity drive and see the change in our oscillation rate. 
A spin 1/2 system is essentially a two-level qubit and a detuning on the cavity drive creates a chevron pattern as shown in Fig.~\ref{fig:sf_chevron}(a,e). 
For spin 1, the oscillation rate and population in cavity also change with the detuning. 
Simulations of our system match experimental data as shown in Fig.~\ref{fig:sf_chevron}(b-d,f-h).
These chevron plots are a necessary step for calibrating the correct frequency for resonant cavity drive in presence of frequency comb on cavity and qubit.
We found that in the presence of two-cavity drives the frequency at which the population of $P_0$ is maximally reduced and $P_{2J}$ is maximally increased for spin $J$ is about \SI{9}{\kilo\hertz} less than the cavity frequency calibrated using cavity state revival technique \cite{kevin_chou_thesis}.
This detuning is observed to be the same for all spins in experiment, so we conclude that this originates from having two cavity drives and not a stark shift because of having multi-frequency qubit drive. 

\begin{figure*}
\centering
\includegraphics[width = \textwidth]{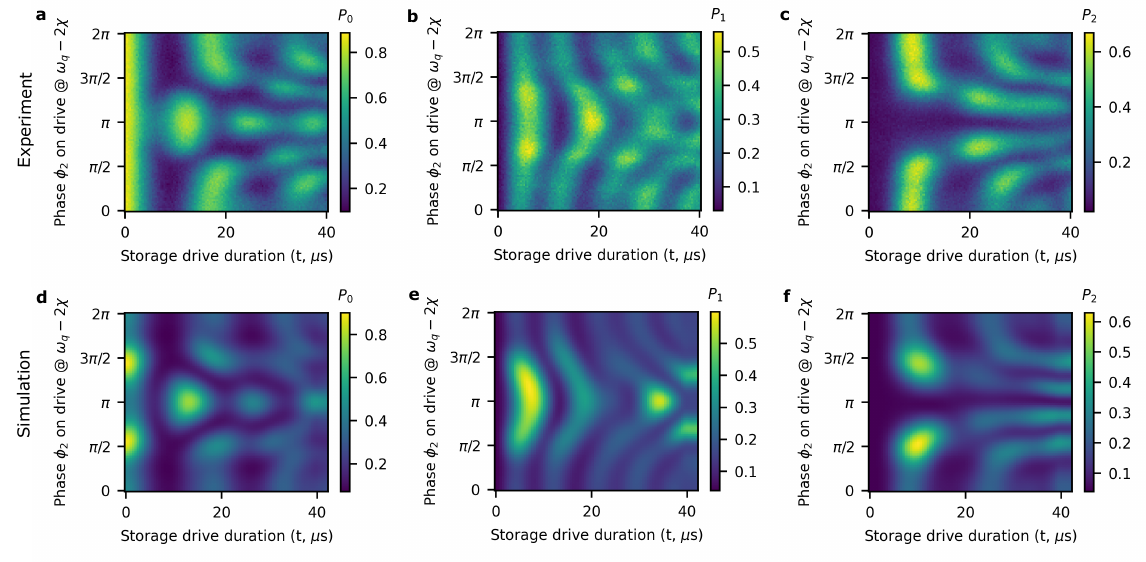}
\caption{\textbf{Spin 1 phase variation:} Phase $\phi_2$ on the qubit drive at frequency $\omega_q + 2\chi$ is varied from 0 to 2$\pi$ and the population in cavity is probed. As phase is changed, the oscillation period and shape changes. Particularly at phase $\pi$, there is no population in $N=2$ cavity state. Hence, the spin 1 system is reduced to a spin 1/2 system. 
}
\label{fig:sf_spin1_phase}
\end{figure*}

\subsection{Qubit drive phase variation for spin 1}

\label{app:spin1_phase_var}

Here we vary the phase on the qubit drive at frequency $f_q + 2\chi$ and check the dynamics of cavity population of the Fock states $|0\rangle, |1\rangle, |2\rangle$. At phase = $\pi$, the matrix element for transition from $|1\rangle$ to $|2\rangle$ is set to zero, and hence the system is reduced to a spin 1/2 system where we only see oscillations between $|0\rangle$ and $|1\rangle$ .

\balancecolsandclearpage

\section{Spin Wigner Function}

\subsection{Theoretical construction}

\label{app:spinwig_theory}

We show in this section that the spin rotations demonstrated above can be used to measure the spin Wigner function.
For a standard harmonic oscillator, the Wigner function associates to each state $\rho$ a function in phase space $(q,p)$ that can be computed from
\begin{equation}
\begin{aligned}
    W_\rho(q,p) &= \mathrm{Tr}\left[\hat \Delta(q,p)\rho\right],\\
    \hat \Delta(q,p) &= 2\hat D(q,p)\expo{i \pi \had \ha}\hat D^\dagger(q,p),
\end{aligned}
\end{equation}
where we have defined $\hat D(q,p) = \exp(i \hat x p - i \hat p q)$. Measuring the Wigner function amounts to displacing the state by $(-q,-p)$ and then measuring the photon number parity. The definition above differs from the usual definition by a factor $1/2\pi$ in order to correspond to the $J\rightarrow \infty$ limit of the spin Wigner function below. The bounds of our spin Wigner function definition is shown in fig. \ref{app:spinwig_theory}.

For a spin state, a spin Wigner function can be defined analogously, where the linear displacements are replaced by spin rotations $\hat R_J(\theta,\phi) = \expo{i \theta (\cos\phi \hat{J}^{(J)}_x + \sin\phi \hat{J}^{(j)}_y)}$ and the parity operator by
\begin{equation}
\begin{aligned}
    \hat \Delta(\theta,\phi) &= \hat R_J(\theta,\phi)\hat \Delta \hat R_J^\dagger(\theta,\phi),\\
    \hat \Delta &= \sum_{m=-J}^J \sum_{l=0}^{2J}\frac{2l+1}{\sqrt{2J(2J+1)}} C^{J m}_{Jm,l0}\ketbra{m+J},\\
    &= \sum_{m=-J}^J \Delta_m \ketbra{m+J},
    \end{aligned}
\end{equation}
where the $C^{J M}_{j_1 m_1 j_2 m_2}$are the Clebsch-Gordan coefficients. The spin Wigner function is correspondingly defined as~\cite{stratonovich_distributions_1957,agarwal_relation_1981,koczor_continuous_2020}
\begin{equation}
\begin{aligned}
    W^{(J)}_\rho(\theta,\phi) &= \mathrm{Tr}\left[\hat \Delta(\theta,\phi)\rho\right]~.\\
\end{aligned}
\end{equation}
The map $W: O \rightarrow W_O^{(J)}$ respects the Stratonovich-Weyl postulates
\begin{enumerate}
    \item Linearity: $W$ is linear and one-to-one
    \item Reality: 
    \begin{equation}
        W_{O^\dag}^{(J)} = \left(W_{O}^{(J)}\right)^*~.
    \end{equation}
    \item Standardization: Defining $d\Omega = \sin\theta d\theta d\phi$, we have
    \begin{equation}
        \mathrm{Tr}(O) = \frac{2J+1}{4\pi} \int_{J^2} d\Omega\, W_{O}^{(J)}(\theta,\phi)~.
    \end{equation}
    \item Traciality: for two operators $O$ and $P$
    \begin{equation}
        \mathrm{Tr}(O M) = \frac{2J+1}{4\pi} \int_{J^2} d\Omega\, W_{O}^{(J)}(\theta,\phi) W_{M}^{(J)}(\theta,\phi)~.
    \end{equation}
    \item Covariance:
    \begin{equation}
        W^{(J)}_{\hat R_S(\alpha,\beta) O \hat R^\dag_S(\alpha,\beta)}(\theta,\phi) = W_O^{(J)}(\theta',\phi'),
    \end{equation}
    where $\vec n' = (\sin\theta' \cos\phi',\sin\theta'\sin\phi',\cos\theta') = \left[R(\hat R_J(\alpha,\beta))\right]^{-1}\cdot (\sin\theta \cos\phi,\sin\theta\sin\phi,\cos\theta )$ with $R(U)$ the $SO(3)$ representation of the $SU(2)$ element $\hat R_J(\alpha,\beta)$.
\end{enumerate}

We can measure this spin Wigner using SU(2) spin rotations and Fock-dependent qubit rotations. 
Let us redefine the Wigner function as
\begin{equation}
\begin{aligned}
    W_\rho(\theta,\phi) &= \mathrm{Tr}\left[\hat \Delta \rho_{\theta,\phi}\right]\\
    \rho_{\theta,\phi} &= \hat R_J^\dag(\theta,\phi) \rho \hat R_J(\theta,\phi) ~.
\end{aligned}
\end{equation}
We can rewrite the matrix elements of $\hat \Delta$ with
\begin{equation}
\begin{aligned}
    \gamma_m &= \frac{\Delta_m - \Delta_{\mathrm{min}}}{\Delta_{\mathrm{max}} - \Delta_{\mathrm{min}}}\\
    \hat \gamma &= \sum_m \gamma_m  \ketbra{m},\\
     \hat \Delta   &= \Delta_{\mathrm{min}} + (\Delta_{\mathrm{max}} - \Delta_{\mathrm{min}})\hat \gamma,
\end{aligned}
\end{equation}
where we have defined $\Delta_{\mathrm{min}} = \mathrm{min}(\{\Delta_m\})$ and $\Delta_{\mathrm{max}} = \mathrm{max}(\{\Delta_m\})$
Here, importantly, the eigenvalues of $\hat \gamma$ are given by $\gamma_m \in [0,1]$. 
We can thus define a set of angles $\theta_m$ such that
\begin{equation}
    \cos^2\left(\frac{\theta_m}{2}\right) = \gamma_m ~.
\label{eqn:swig_theta_1}
\end{equation}
We can measure the Wigner function through a qubit measurement after the pulse
\begin{equation}
    \hat U_\gamma = \sum_m \hat R_x(\theta_m)\otimes \ketbra{m}~,
\end{equation}
which rotates the qubit by a different angle for each Fock state $\ket{m}$, where the angle depends on $\gamma_m$.
The probability of measuring $\ket{g}$ after the (inverse) SU(2) rotation and the pulse $\hat U_\gamma$, assuming that the qubit is initialized in $\ket{g}$, is
\begin{equation}
\begin{aligned}
    \langle \ketbra{g}\otimes \mathbb I \rangle &= \mathrm{Tr}\left[\hat U_\gamma (\ketbra{g}\otimes\rho_{\theta,\phi})\hat U^\dag_\gamma (\ketbra{g} \otimes \mathbb I) \right] \\
    &= \sum_m |\bra{e}\hat U_\gamma \ket{g}|^2 \mathrm{Tr}[\rho_{\theta,\phi} \ketbra{m}]\\
    &= \sum_m \cos^2\left(\frac{\theta_m}{2}\right) \mathrm{Tr}[\rho_{\theta,\phi} \ketbra{m}]\\
    &= \langle \hat \gamma \rangle ~.\\
\end{aligned}
\end{equation}
Rescaling this probability by $(\gamma_{\mathrm{max}} - \gamma_{\mathrm{min}})$ and adding $\gamma_{\mathrm{min}}$ gives us $W_\rho(\theta,\phi)$. 

In order to symmetrize the effect of readout errors, the expectation value of $\hat \Delta$ can also be decomposed as 
\begin{equation}
\begin{aligned}
    \eta_m &= \frac{\Delta_{\mathrm{max}} - \Delta_m}{\Delta_{\mathrm{max}} - \Delta_{\mathrm{min}}}\\
    \hat \eta &= \sum_m \eta_m  \ketbra{m},\\
     \hat \Delta   &= \Delta_{\mathrm{max}} - (\Delta_{\mathrm{max}} - \Delta_{\mathrm{min}})\hat \eta.
\end{aligned}
\end{equation}
Accordingly, we can measure this operator by defining 
\begin{equation}
    \cos^2\left(\frac{\beta_m}{2}\right) = \eta_m ~,
\label{eqn:swig_theta_2}
\end{equation}
and we can measure the Wigner function by measuring the qubit population in the $\ket{g}$ state after the pulse
\begin{equation}
    \hat U_\eta = \sum_m \hat R_x(\beta_m)\otimes \ketbra{m}~.
\end{equation}

\begin{figure*}
\centering
\includegraphics[width = 0.5\textwidth]{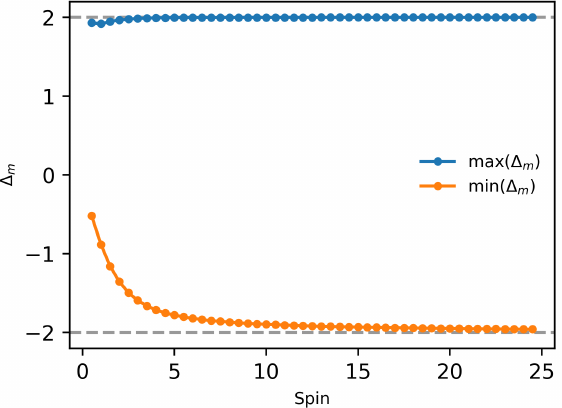}
\caption{\textbf{Limits of spin Wigner function:} The theoretically possible maximum and minimum values of spin Wigner function as a function of the size of the spin. 
As the spin size becomes large (or tends to infinite), the maximum and minimum values tend to $\pm 2$. 
}
\label{fig:sf_spinwig_theory}
\end{figure*}

\subsection{Experimental implementation}

\begin{figure*}
\centering
\includegraphics[width = \textwidth]{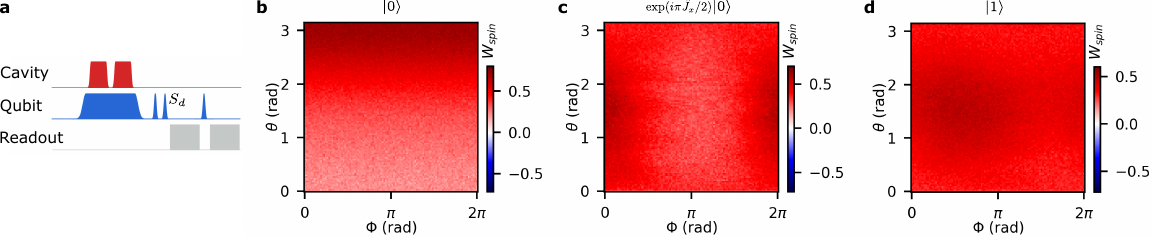}
\caption{\textbf{Spin Wigner:} (a) Pulse schematic for measuring the spin Wigner function for Spin 1. (b-d) Spin Wigner functions for (b) Fock state $|0\rangle$, (c) spin coherent state on the equator, (d) Fock state $|1\rangle$, which is also a equatorial spin cat state. 
Due to infidelities of spin rotations and measurements, the contrast of spin Wigner function is poor. 
So we had to scale the colorbar differently for better visibility in (b-d).   
}
\label{fig:sf_spinwigner}
\end{figure*}

In this section we demonstrate that the spin Wigner functions can be experimentally measured as described in App.~\ref{app:spinwig_theory}. 
The schematic pulse sequence is shown in Fig.~\ref{fig:sf_spinwigner}. 
First, we choose a qubit drive duration for the experiment based on App.~\ref{app:technical_calibration} and Fig.~\ref{fig:sf_background}.
The first cavity drive is used to prepare the state we want to probe.
For Fock state $|0\rangle$ and $|1\rangle$ we do not need this first drive as cavity $|0\rangle$ is prepared by having an idle period much longer than the cavity lifetime ($T_1$) and $|1\rangle$ is prepared using a conventional photon blockade procedure. 
The second cavity drive is used to accomplish a spin rotation for angle $\theta$ around an axis with angle $\phi$ with respect to the X axis. 
As we do for the spin cat gates, a $\phi$ rotation is accomplished by a frame rotation of the cavity drive and a $\theta$ rotation is achieved by varying the duration of the cavity drive.
After the qubit drive ends we disentangle the qubit from cavity using SNAP gate that consists two consecutive number-selective $\pi$ pulses on the qubit for a total duration of \SI{2}{\micro\second}. 
Then we readout the qubit state and store it for post-processing. 
Finally, we do number-selective qubit rotations and readout the final state of the qubit. 
The probability of finding the qubit in ground state ($P_g$) directly relates to the spin Wigner function. 

For comparison to bosonic Wigner measurements, the spin rotation is analogous to displacement operation and the number-selective qubit rotations are analogous to the parity measurement.
For symmetrizing errors during spin rotations, we do two different number-selective qubit rotations.
For the first case, we do the $\theta_m$ rotations on the qubit following Eq.~\ref{eqn:swig_theta_1} and define the spin Wigner function as 
\begin{equation}
W^{(1)}_\textrm{spin}(\theta, \phi)= \Delta_\textrm{min} + P_g (\Delta_\textrm{max} - \Delta_\textrm{min})
\end{equation}
In a second experiment, we do $\beta_m$ rotations on the qubit following Eq.~\ref{eqn:swig_theta_2} and define the spin Wigner as 
\begin{equation}
W^{(2)}_\textrm{spin}(\theta, \phi)= \Delta_\textrm{max} - P_g (\Delta_\textrm{max} - \Delta_\textrm{min})
\end{equation}
Finally, in a post-processing we select the cases where our first readout measured the qubit in ground state (implying the spin rotation was successful) and then average the two spin Wigner functions measured above to get the error-symmetrized spin Wigner function.
\begin{equation}
W_\textrm{spin}(\theta, \phi) = (W^{(1)}_\textrm{spin}(\theta, \phi) +  W^{(2)}_\textrm{spin}(\theta, \phi))/2
\end{equation}
Spin Wigner functions measured this way are shown in Fig.~\ref{fig:sf_spinwigner} for spin 1:  Fock state $|0\rangle$, $|1\rangle$, and a spin coherent state on the equator of spin 1 Bloch sphere. 
Since our spin rotations suffer from errors, the measured spin Wigner functions are noisy and imperfect, with poor contrast. 

\bibliography{sr-bibliography,vf-bib}

\end{document}